\documentclass[12pt,a4paper]{article}

\usepackage{amsmath}
\usepackage{amssymb}
\usepackage{graphics}
\usepackage{graphicx}
\usepackage{epsfig}
\usepackage{dsfont}
\usepackage{overpic}

\let\oldappendix=\appendix
\let\oldsection=\section
\renewcommand{\appendix}{\oldappendix%
\def\theequation{\Alph{section}.\arabic{equation}}%
\renewcommand{\section}{\setcounter{equation}{0}\oldsection}}

\newcommand{\beq}{\begin{equation}}
\newcommand{\eeq}{\end{equation}}
\newcommand{\beqa}{\begin{eqnarray}}
\newcommand{\eeqa}{\end{eqnarray}}
\newcommand{\no}{\nonumber}
\newcommand{\q}{\quad}
\newcommand{\qq}{\qquad}

\newcommand{\tr}{\mbox{tr}}
\newcommand{\sfrac}[2]{{\textstyle\frac{#1}{#2}}}

\newcommand{\deltaph}{\lambda}
\def\bra#1{\left\langle #1\right|}
\def\ket#1{\left| #1\right\rangle}
\def\trf#1{\left\langle #1 \right\rangle}

\newcommand{\newop}[2]{\def#1{\mathop{\mathrm{#2}}\nolimits}}
\newop{\artanh}{artanh}
\newop{\det}{det}
\newop{\tr}{tr}
\newop{\diag}{diag}

\newcommand{\tad}[1]{\Delta_{#1}}

\newcommand{\cder}{D}
\newcommand{\decay}{f}
\newcommand{\coeffv}[2]{v_{#1}^{(#2)}}
\newcommand{\cbeta}[2]{\beta_{#1}^{(#2)}}
\newcommand{\cvtwid}[2]{\tilde{v}_{#1}^{(#2)}}

\newcommand{\eV}{\,\mathrm{eV}}

\newcommand{\MeV}{\,\mathrm{MeV}}
\newcommand{\GeV}{\,\mathrm{GeV}}
\newcommand{\Lagr}{\mathcal{L}}

\begin{document}

\hfill 

\hfill 

\bigskip\bigskip

\begin{center}

{{\Large\bf  $\mbox{\boldmath$\eta, \eta' \to \pi^+ \pi^- \gamma$}$
  with coupled channels}}

\end{center}

\vspace{.4in}

\begin{center}
{\large B. Borasoy\footnote{email: borasoy@ph.tum.de},
 R. Ni{\ss}ler\footnote{email: rnissler@ph.tum.de}}

\bigskip

\bigskip

Physik Department\\
Technische Universit{\"a}t M{\"u}nchen\\
D-85747 Garching, Germany \\

\vspace{.2in}

\end{center}

\vspace{.7in}

\thispagestyle{empty} 

\begin{abstract}
The decays $\eta, \eta' \to \pi^+ \pi^- \gamma$ are investigated
within an approach that combines one-loop chiral perturbation theory with
a coupled channel Bethe-Salpeter equation which satisfies unitarity constraints and generates
vector mesons dynamically from composite states of two pseudoscalar mesons.
It is furthermore shown that the inclusion of the $\eta'$ as a dynamical degree 
of freedom does not renormalize the Wess-Zumino-Witten term.

\end{abstract}\bigskip

\begin{center}
\begin{tabular}{ll}
\textbf{PACS:}&12.39.Fe \\[6pt]
\textbf{Keywords:}& Chiral Lagrangians, anomaly, unitarity.
\end{tabular}
\end{center}


\vfill

\section{Introduction}\label{sec:intro}

The decay $\eta \to \pi^+ \pi^- \gamma$ is determined entirely by the chiral anomaly,
if the quark masses of the up, down and strange quarks and the involved four-momenta
are sent to zero. The kinematical region of the decay, on the other hand, is constrained
to $4 m_\pi^2 \le (p^+ + p^-)^2 \le m_\eta^2$ with $p^+$ and $p^-$ the four-momenta
of the $\pi^+$ and $\pi^-$, respectively, and thus far from the zero momentum limit
which is described by the anomalous Wess-Zumino-Witten (WZW) term
in the chiral effective Lagrangian \cite{WZ, W}.

In the framework of chiral effective field theory higher order contact interactions
as well as unitarity corrections which arise from loop graphs will be of importance
and must be included in order to enable a proper description of the experimental
decay width and the photon spectrum, see, {\it e.g.}, \cite{BBC} for a one-loop
calculation and \cite{Gor, Lay}
for experimental results.

For the decay $\eta' \to \pi^+ \pi^- \gamma$ contributions from vector meson exchange
will dominate the amplitude and unitarity effects should be implemented via final
state interactions. This is clearly beyond the perturbative framework of chiral
perturbation theory (ChPT) and requires utilization of non-perturbative tools
which match onto the results from ChPT.
One obvious approach would be to employ the vector dominance picture
with energy-dependent widths in the vector meson propagators. However, this procedure
can be shown to be in contradiction to the one-loop result of ChPT \cite{H}.

Another possibility is to postulate an $N/D$ structure for the decay amplitude
which matches onto both the one-loop chiral corrections and vector meson dominance
in the pertinent limits. This approach has been applied successfully in \cite{VH},
where the decay width and photon spectra of both the $\eta$ and $\eta'$ decay are
brought to agreement with experiment \cite{CB, GAMS} and constraints
for the $\eta$-$\eta'$ mixing angle have been given.

In the present investigation we will apply an alternative approach which relies
solely on chiral symmetry and unitarity and has already been employed successfully 
in the anomalous two-photon decays of $\pi^0, \eta$ and $\eta'$ \cite{BN}.
Therein the one-loop contributions from the WZW Lagrangian were calculated
and unitarity corrections beyond one loop were
included by employing a coupled channel
Bethe-Salpeter equation which satisfies unitarity constraints
and generates vector mesons from composite states of pseudoscalar mesons.
Although being similar in spirit to the complete vector meson dominance 
picture, this approach clearly distinguishes between the exchange of either one
or two vector mesons.
It turns out that for the two-photon decays the exchange of one vector meson is the dominant contribution,
whereas the simultaneous exchange of two vector mesons is suppressed.
This is in contradistinction to complete vector meson dominance where the coupling
of photons to pseudoscalar mesons is always mediated by vector mesons.

The purpose of the present work is to extend this approach to the decays
$\eta , \eta' \to \pi^+ \pi^- \gamma$. We will first perform a complete
one-loop calculation including all counter terms of unnatural parity up to
chiral order $p^6$ and with the $\eta'$ as a dynamical degree of freedom,
but without employing large $N_c$ counting rules. 
As we will show explicitly, the inclusion of the massive $\eta'$ state
does not lead to a renormalization of the WZW Lagrangian and therefore
satisfies constraints from the anomalous Ward identities.
Unitarity corrections and final state interactions are then appended within a
coupled channel analysis which can be easily matched to the ChPT result.

The importance of resonance exchange for these decays can be studied without
including vector mesons explicitly in the effective Lagrangian.
Moreover, we will critically examine the issue of $\eta$-$\eta'$ mixing for these decays.

This work is organized as follows. In the next section, the complete one-loop
calculation of the decays is performed. The inclusion of unitarity corrections
beyond one-loop is outlined in Sec.~\ref{sec:CC} and numerical results are presented in 
Sec.~\ref{sec:num} along with a comparison with experimental data. 
Sec.~\ref{sec:concl} contains our conclusions.

\section{One-loop calculation} \label{sec:1loop}

The decays $\eta, \eta' \to \pi^+ \pi^- \gamma$ arise from the unnatural parity part of the 
effective Lagrangian which collects the terms that are proportional to the tensor
$\epsilon_{\mu \nu \alpha \beta}$.
Within the effective theory the chiral anomalies of the underlying QCD Lagrangian
are accounted for by the WZW term \cite{WZ, W, KL1} \footnote{Note that for our
purposes we can safely set the singlet axial vector field $\langle a_\mu \rangle$
and the derivative of the QCD vacuum angle, $\partial_\mu \theta$, to zero in $S_{\scriptscriptstyle{WZW}}$
which
enables us to work with the renormalization group invariant form of the anomaly.}
\beq  \label{eq:wzw}
S_{\scriptscriptstyle{WZW}} (U,v) = \int d^4 x
\mathcal{L}_{\scriptscriptstyle{WZW}} = - \frac{i }{80 \pi^2} \int_{M_5} \langle \Sigma^5 \rangle
    -  \frac{i}{16 \pi^2} \int_{M_4} W(U,v)
\eeq
where
\beq
W(U,v) = i \trf{U \,d U^\dagger U \,d U^\dagger U \,d U^\dagger v -
                U^\dagger d U \,U^\dagger d U \,U^\dagger d U \,v}
\eeq
with $\Sigma = U^\dagger dU$ and for the number of colors we set $N_c=3$.
The matrix valued field 
$U = \exp\{i \sqrt{2} \phi / \decay\}$ contains the 
Goldstone boson octet ($\pi, K, \eta_8$) and the singlet field $\eta_0$,
where $\decay$ is the pseudoscalar decay constant in the chiral limit.
The expression $\trf{\ldots}$ denotes the trace in flavor space and 
we have displayed only the pieces of the Lagrangian relevant for the present work.
We utilized, furthermore, the differential form notation
\beq
v = dx^\mu v_\mu , \qq d = dx^\mu \partial_\mu
\eeq
with the Grassmann variables $dx^{\mu}$ which yield the volume element $dx^\mu
dx^\nu dx^\alpha dx^\beta = \epsilon^{\mu \nu \alpha \beta} d^4x$.
In the second integral Minkowskian space $M_4$ is extended to a five-dimensional
manifold $M_5$ and the $U$ fields are functions on $M_5$, {\it cf.} \cite{W, KL1} for details.
The external vector field $v= -e Q A$ describes the coupling of the photon field
$A=dx^\mu A_\mu$ to the mesons with $Q= \frac{1}{3} \mbox{diag}(2,-1,-1)$ being the charge
matrix of the light quarks.

In addition to the WZW anomaly action there is one more term of unnatural parity 
at fourth chiral order 
contributing to $\eta' \to \pi^+ \pi^- \gamma$ at leading order
and to the $\eta$ decay via mixing at next-to-leading order ${\cal O}(p^6)$.
This gauge invariant term arises due to the extension to the $U(3)$ framework and reads 
\beq \label{eq:unnatp}
d^4x \Lagr_{ct}^{(4)} = i \,W_3
\trf{d U \,d U^\dagger d v + d U^\dagger d U \,d v} \ ,
\eeq
where $W_3$ is a function of $\eta_0$,
$W_3(\eta_0/\decay)$. The potential $W_3$ 
can be expanded in the singlet field with coefficients $w_{3}^{(j)}$ that are 
not fixed by chiral symmetry, while 
parity conservation implies that $W_3$ is an odd function of $\eta_0$.

\begin{figure}
\centering
\includegraphics[scale=1.0]{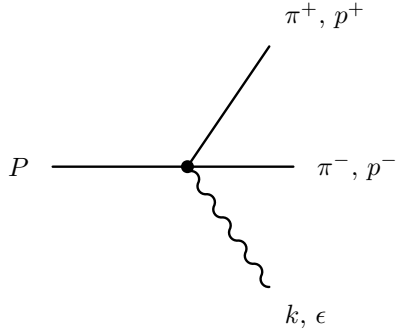}
\caption{Tree diagram of the decay $P \to \pi^+ \pi^- \gamma$,
         where $p^+$ and $p^-$ denote the momenta of the outgoing pions.
         The momentum and the polarization of the photon are indicated by 
         $k$ and $\epsilon$, respectively.}
\label{fig:tree}
\end{figure}

At the tree level the contributing diagram is depicted in Fig.~\ref{fig:tree} with $P$ 
symbolizing either an $\eta$ or $\eta'$. 
Expanding $U$ in Eqs.~(\ref{eq:wzw}) and~(\ref{eq:unnatp}) in terms of $\phi$ yields 
the tree level vertex
\beqa \label{eq:vtree}
d^4x \Lagr_{\textrm{WZW}} & = & \frac{i \sqrt{2}}{4 \pi^2 \decay^3} 
\trf{d \phi \,d \phi \,d \phi \,v} + \ldots \, , \no \\
d^4x \Lagr_{ct}^{(4)} & = & w_{3}^{(1)} \frac{4 i}{f^3} \eta_0 
\trf{d \phi \,d \phi \,d v} + \ldots \ .
\eeqa
From these terms we can derive the tree level amplitude
\beq \label{eq:Atree}
\mathcal{A}^{\textit{(tree)}} (P \rightarrow \pi^+ \pi^- \gamma) =
  - e k_\mu \epsilon_\nu p^{+}_\alpha p^{-}_\beta
  \epsilon^{\mu \nu \alpha \beta} \frac{1}{4 \pi^2 \decay^3} \alpha_{P}^{\textit{(tree)}}
\eeq
with 
\beq
\alpha_{\eta}^{\textit{(tree)}}  = \frac{1}{\sqrt{3}} \,, \qq \qq
\alpha_{\eta'}^{\textit{(tree)}} = \sqrt{\frac{2}{3}} - 16 \pi^2 w_{3}^{(1)} \ ,
\eeq
and $w_{3}^{(1)}$ is the coefficient of the leading term in $W_3$.

\subsection{One-loop diagrams}

\begin{figure} [bh]
\centering
\includegraphics[scale=1.0]{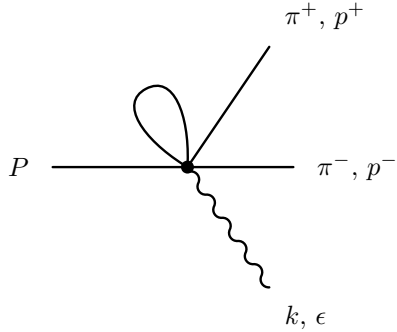}
\caption{Tadpole diagram which contributes to $P \to \pi^+ \pi^- \gamma$.}
\label{fig:tad}
\end{figure}

There are four different topologies of one-loop graphs contributing to the $\pi^+ \pi^- \gamma$ decays. 
We start with the discussion of the tadpole diagram shown in Fig.~\ref{fig:tad}.
The pertinent terms from both the WZW and the unnatural parity Lagrangian are
{\arraycolsep3pt \beqa   \label{eq:vertad}
d^4x \Lagr_{\textrm{WZW}} & = & -\frac{i \sqrt{2}}{24 \pi^2 f^5}
\big\langle(\sfrac{1}{2} \phi [\phi \,, d\phi][d\phi \,,d\phi] 
    + d\phi[d\phi \,,\phi[\phi \,, d\phi]] \no \\
& & \qq \qq \qq + [[d\phi \,,d\phi] d\phi \,,\phi^2] 
    + \sfrac{3}{2} \phi[\phi \,,[d\phi \,,d\phi] d\phi])\,v \big\rangle + \dots \, , \no \\
d^4x \Lagr_{ct}^{(4)} & = & w_{3}^{(1)} \frac{2 i}{3 f^5} \eta_0
\trf{( [\phi \,,d\phi][\phi \,,d\phi] - [\phi \,,[\phi \,,[d\phi \,,d\phi]]]) \,d v} 
\,, \no \\
& & + \ w_{3}^{(3)} \frac{4 i}{f^5} \eta_{0}^3 \trf{d\phi \,d\phi \,dv} + \dots \ .
\eeqa}%
The $w_{3}^{(3)}$-term yields an $\eta'$ tadpole which---strictly speaking---spoils the 
chiral counting scheme, as it contributes at fourth chiral order. 
However, the $\eta'$ tadpole does not 
contain any infrared physics and can be absorbed completely into the low-energy constants 
(LECs) of the effective Lagrangian, as it is neither a function of the Goldstone boson masses 
nor of the external momenta, it is just a constant.
This is consistent with the fact that in infrared regularization
the $\eta'$ tadpole vanishes, whereas the tadpoles for the 
Goldstone boson octet remain unaltered. 
By defining
\beq \label{eq:rentad}
w_{3}^{(1)} = w_{3}^{(1)r} - w_{3}^{(3)} \Delta_{\eta'} / \decay^2
\eeq
the $\eta'$ tadpole is compensated by the second term in Eq.~(\ref{eq:rentad}) and the chiral counting 
scheme is restored without renormalizing the WZW Lagrangian.
The expression $\Delta_\phi$ symbolizes the finite part of the tadpole integral
\beq
\Delta_\phi = \left( \int \frac{d^d l}{(2 \pi)^d} \frac{i}{l^2 - m_\phi^2 +i\varepsilon}
              \right)_{\textit{finite}}
 = \frac{m_\phi^2}{16 \pi^2} \,  \ln \frac{m_\phi^2}{\mu^2} 
\eeq
with $\mu$ being the scale introduced in dimensional regularization.
In the present work we are only concerned with the finite pieces of the diagrams
and neglect the divergent portions throughout.
Note that there are further contributions from one-loop graphs and counter terms which require
a renormalization of $w_{3}^{(1)}$ as will be discussed below.

The remaining contributions in Eq.~(\ref{eq:vertad}) to the amplitudes
involve only Goldstone boson tadpoles  and 
are thus of next-to-leading order
\beq  \label{eq:Atad}
\mathcal{A}^{\textit{(tad)}} (P \rightarrow \pi^+ \pi^- \gamma) =
  - e k_\mu \epsilon_\nu p^{+}_\alpha p^{-}_\beta
  \epsilon^{\mu \nu \alpha \beta } \dfrac{1}{4 \pi^2 f^5} \Big(
  \beta_P^{\textit{(tad)},\pi} \Delta_\pi 
  + \beta_P^{\textit{(tad)},K} \Delta_K \Big)
\eeq
with 
\beq
\begin{array}{lcllcl}
\beta_{\eta}^{\textit{(tad)},\pi}  & = & -\dfrac{5}{3 \sqrt{3}} \,, \qq &
\beta_{\eta'}^{\textit{(tad)},\pi} & = & -\dfrac{5}{3} \left( \sqrt{\dfrac{2}{3}}
                                                   - 16 \pi^2 w_{3}^{(1)r} \right) \,, \\
\beta_{\eta}^{\textit{(tad)},K}    & = & -\dfrac{4}{3 \sqrt{3}} \,, \qq &
\beta_{\eta'}^{\textit{(tad)},K}   & = & -\dfrac{5}{6} \left( \sqrt{\dfrac{2}{3}}
                                                   - 16 \pi^2 w_{3}^{(1)r} \right) \,.
\end{array}
\eeq
Note that we have replaced the coefficient $w_{3}^{(1)}$ by the renormalized coupling 
$w_{3}^{(1)r}$ which is consistent at the one-loop level.

\begin{figure}
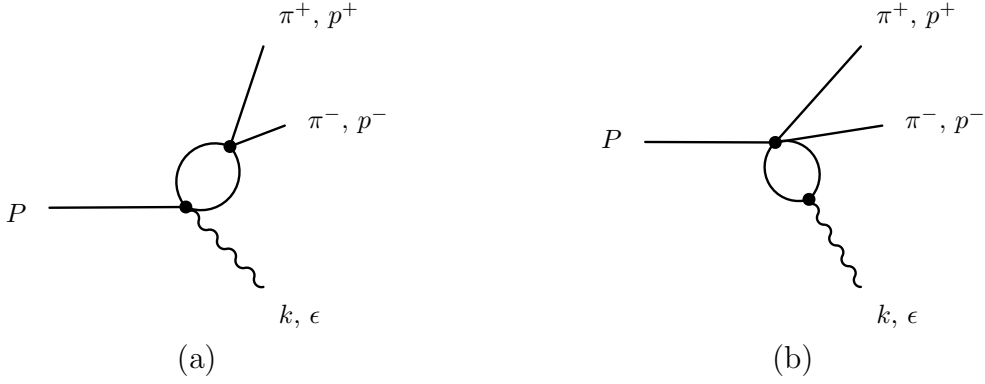

\centering
\begin{minipage}[b]{0.35\textwidth}
\centering
\includegraphics[scale=1.0]{feynps.15} \\
(a)
\end{minipage}
\hspace{0.1\textwidth}
\begin{minipage}[b]{0.35\textwidth}
\centering
\includegraphics[scale=1.0]{feynps.16} \\
(b)
\end{minipage}
\caption{One-loop diagrams contributing to $P \to \pi^+ \pi^- \gamma$.}
\label{fig:1la}
\end{figure}

The loop graph in Fig.~\ref{fig:1la}a has a four-meson vertex
from the lowest order Lagrangian $\Lagr^{(0+2)}$ of natural parity
\beqa
\Lagr^{(0+2)} & = & \frac{f^2}{4} \trf{D_\mu U^\dagger D^\mu U} + \ldots 
                =   \frac{1}{12 \decay^2}  \trf{[\phi \,,\partial_\mu \phi][\phi \,,\partial^\mu \phi]} 
		    + \ldots \,, \no \\
D_\mu U & = & \partial_\mu U + i [U , v_\mu] \ ,
\eeqa
which yields the amplitudes
\beq  \label{eq:A1la}
\mathcal{A}^{\textit{(a)}} (P \rightarrow \pi^+ \pi^- \gamma) =
  - e k_\mu \epsilon_\nu p^{+}_\alpha p^{-}_\beta
  \epsilon^{\mu \nu \alpha \beta } \dfrac{1}{4 \pi^2 f^5} 
  \Big(  \beta_P^{\textit{(a)},\pi} I_1(m_\pi^2;s_{+-})
       + \beta_P^{\textit{(a)},K}   I_1(m_K^2;s_{+-})  \Big) 
\eeq
with $s_{+ -} = (p^+ + p^-)^2$. The integral $I_1$ is defined by
\beq \label{eq:intI1}
I_1(m^2;p^2) =   \frac{2 }{3} \Big(  \frac{1}{2} \Delta  +  (m^2 - \frac{p^2}{4}) G_{mm}(p^2)
      + \frac{1}{96 \pi^2} (p^2 -6 m^2)     \Big) ,
\eeq
where $G$ is the finite part of the scalar one-loop integral
\beqa \label{eq:intG}
G_{m \bar{m}}(p^2) & = & \left( \int\frac{\,d^d l}{(2\pi)^d}\,
\frac{i}{(l^2-m^2+i \epsilon)( (l-p)^2-\bar m^2+i \epsilon)} \right)_{\textit{finite}}
\no \\[2ex]
& = & \frac{1}{16\pi^2}\bigg[-1+ \ln\frac{m \bar{m}}{\mu^2}
      +\frac{m^2-\bar{m}^2}{p^2}\ln\frac{m}{\bar{m}} \bigg. \no \\[2ex]
& & \qq \qq  \bigg. -\frac{2\sqrt{\deltaph_{m\bar{m}}(p^2)}}{p^2}\artanh
    \frac{ \sqrt{\deltaph_{m\bar{m}}(p^2)}}{(m+\bar{m})^2-p^2} \bigg] , \\[3ex]
\deltaph_{m\bar{m}}(p^2) & = & \big((m-\bar{m})^2-p^2\big)\big((m+\bar{m})^2-p^2\big) \no .
\eeqa
The coefficients $\beta_P^{\textit{(a)},\phi}$ read
\beq
\begin{array}{lcllcl}
\beta_{\eta}^{\textit{(a)},\pi} & = & \dfrac{1}{\sqrt{3}} \,, \qq &
\beta_{\eta'}^{\textit{(a)},\pi} & = & \sqrt{\dfrac{2}{3}} - 16 \pi^2 w_{3}^{(1)r} 
\,, \\[3ex]
\beta_{\eta}^{\textit{(a)},K} & = & \dfrac{2}{\sqrt{3}} \,, \qq &
\beta_{\eta'}^{\textit{(a)},K} & = & \dfrac{1}{\sqrt{6}} - 8 \pi^2 w_{3}^{(1)r} \,.
\end{array}
\eeq

The diagram in Fig.~\ref{fig:1la}b, on the other hand, 
involves the five-meson vertex of the WZW action
\beq
\frac{-i}{80 \pi^2} \int_{x^5=0}^{x^5=1} \trf{\Sigma^5} 
= \frac{\sqrt{2}}{20 \pi^2 f^5} \trf{\phi \,d\phi \,d\phi \,d\phi \,d\phi} + \ldots \ .
\eeq
The singlet field $\eta_0$ does not contribute to the five-meson vertex and the 
$\eta'$ is only involved via $\eta$-$\eta'$ mixing which is of higher order in this framework. 
At $\mathcal{O}(p^6)$ the loop diagram 
in Fig.~\ref{fig:1la}b thus yields a contribution only to the $\eta$ decay
\beq  \label{eq:A1lb}
\mathcal{A}^{\textit{(b)}} (P \rightarrow \pi^+ \pi^- \gamma) =
  - e k_\mu \epsilon_\nu p^{+}_\alpha p^{-}_\beta
  \epsilon^{\mu \nu \alpha \beta } \dfrac{1}{4 \pi^2 f^5} 
  \beta_P^{\textit{(b)}} \Delta_K
\eeq
with
\beq
\begin{array}{rclrcl}
\beta_{\eta}^{\textit{(b)}}  & = & \sqrt{3} \,, \qq
\beta_{\eta'}^{\textit{(b)}} & = & 0 \ .
\end{array}
\eeq

\begin{figure}
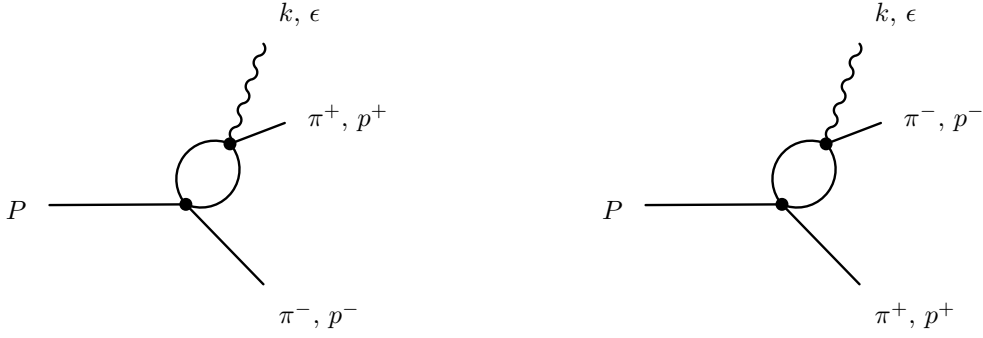

\centering
\begin{minipage}[b]{0.35\textwidth}
\centering
\includegraphics[scale=1.0]{feynps.17}
\end{minipage}
\hspace{0.1\textwidth}
\begin{minipage}[b]{0.35\textwidth}
\centering
\includegraphics[scale=1.0]{feynps.18}
\end{minipage}
\caption{One-loop diagrams with an $\eta' \pi$ loop.}
\label{fig:1letap}
\end{figure}
Two additional one-loop graphs which contribute at sixth chiral order are
depicted in Fig.~\ref{fig:1letap} and an analysis of the $\mathcal{O}(p^2)$ four-meson vertex 
shows that the only contribution to this diagram stems from
\beq
\Lagr^{(0+2)} = V_1 \trf{D_\mu U^\dagger D^\mu U} + \ldots 
= v_{1}^{(2)} \frac{2}{f^4} \eta_{0}^2 \trf{\partial_\mu \phi \partial^\mu \phi} + \ldots \ 
\eeq 
where $V_1(\eta_0 / f) = \decay^2 / 4 + v_1^{(2)} \eta_0^2/\decay^2 + \ldots $ .
The fourth order vertex is given by Eq.~(\ref{eq:vtree}) and we can readily calculate the 
amplitude
\beq \label{eq:Ac}
\mathcal{A}^{\textit{(c)}} (P \rightarrow \pi^+ \pi^- \gamma) = 
  - e k_\mu \epsilon_\nu p^{+}_\alpha p^{-}_\beta
  \epsilon^{\mu \nu \alpha \beta } \dfrac{1}{4 \pi^2 f^5} \ \beta_P^{\textit{(c)}} 
  \big( I_2(m_{\pi}^2, m_{\eta'}^2; s_{+ \gamma}) 
      + I_2(m_{\pi}^2, m_{\eta'}^2; s_{- \gamma})\big)
\eeq
with $s_{+ \gamma} = (p^+ + k)^2$ and $s_{- \gamma} = (p^- + k)^2$ and
\beq
\beta_{\eta}^{\textit{(c)}} = 0 \,, \qq
\beta_{\eta'}^{\textit{(c)}} = 
  -\frac{4 v_{1}^{(2)}}{f^2} \left(\sqrt{\frac{2}{3}} - 16 \pi^2 w_{3}^{(1)r} \right) \,,
\eeq
where we have replaced again $w_{3}^{(1)}$ by $w_{3}^{(1)r}$ and neglected 
two-loop corrections.
The finite part of the $\pi \eta'$ loop integral $I_2$ is given by
\beq \label{eq:intI2}
\begin{split}
I_2(m_{\pi}^2, m_{\eta'}^2; p^2) =
&  \frac{1}{6 p^2}\big\{-(p^2 - (m_{\eta'}-m_{\pi})^2)(p^2 - (m_{\eta'}+m_{\pi})^2)
  G_{m_{\pi} m_{\eta'}}(p^2) \\
& \qquad \; + (p^2 - m_{\eta'}^2 + m_{\pi}^2) \Delta_{\pi}
  + (p^2 + m_{\eta'}^2 - m_{\pi}^2) \Delta_{\eta'} \big\} \\
& + \frac{1}{144 \pi^2}(p^2 - 3 m_{\pi}^2 - 3 m_{\eta'}^2) \, .
\end{split}
\eeq
In order to maintain the matching between the chiral and the loop 
expansion of the amplitude,
we have to absorb the contributions of $\mathcal{O}(p^4)$ in $\mathcal{A}^{\textit{(c)}}$ into the leading order 
coupling $w_{3}^{(1)}$, Eq.~(\ref{eq:vtree}). Expanding $I_2$ in terms of $\frac{p^2}{m_{\eta'}^2}$ and 
$\frac{m_{\pi}^2}{m_{\eta'}^2}$ yields
\beq
I_2(m_{\pi}^2, m_{\eta'}^2; p^2) = \frac{1}{2} \Bigl( \Delta_{\eta'} - \frac{m_{\eta'}^2}{32 \pi^2} \Bigr)
  \Bigl( 1 + \frac{m_{\pi}^2}{m_{\eta'}^2} \Bigr) 
  - \frac{1}{6} \Bigl( \Delta_{\eta'} - \frac{m_{\eta'}^2}{96 \pi^2} \Bigr) \frac{p^2}{m_{\eta'}^2} + \ldots
\eeq
and the renormalization may be accomplished by modifying Eq.~(\ref{eq:rentad}) according to
\beq  \label{eq:ren}
w_{3}^{(1)} = w_{3}^{(1)r} - w_{3}^{(3)} \frac{\Delta_{\eta'}}{f^2}
  - \frac{v_{1}^{(2)}}{4 \pi^2 f^4} 
  \left(\sqrt{\frac{2}{3}} - 16 \pi^2 w_{3}^{(1)r} \right) 
  \Bigl( \Delta_{\eta'} - \frac{m_{\eta'}^2}{32 \pi^2} \Bigr) \ ,
\eeq
where we replaced $w_{3}^{(1)}$ by $w_{3}^{(1)r}$ in the last term of Eq.~(\ref{eq:ren}).
After this renormalization, we work with the subtracted integral
\beq  \label{eq:intI2p}
\hat{I}_2 (m_{\pi}^2, m_{\eta'}^2; p^2) = I_2(m_{\pi}^2, m_{\eta'}^2; p^2) -
   I_2 (0, m_{\eta'}^2; 0)
\eeq
which yields contributions to the amplitude of order $p^6$ and higher. 
Thus all one-loop contributions of $\mathcal{O}(p^4)$ can be absorbed into a 
redefinition of the low-energy constant $w_{3}^{(1)}$ and do not renormalize the WZW
Lagrangian.

We now turn to the next-to-leading order chiral corrections 
from the decay constants, $\eta$-$\eta'$ mixing and 
wave-function renormalization which are readily calculated from the formulae given in 
\cite{BB1}.
Replacing in the tree level result, Eq.~(\ref{eq:Atree}), 
the decay constant in the chiral limit by the corresponding one-loop expression
induces the following $\mathcal{O}(p^6)$ corrections
to the decay amplitudes
\beq \label{eq:Af}
\mathcal{A}^{\textit{(f)}} (P \rightarrow \pi^+ \pi^- \gamma) =
  - e k_\mu \epsilon_\nu p^{+}_\alpha p^{-}_\beta
  \epsilon^{\mu \nu \alpha \beta } \dfrac{1}{4 \pi^2 F_P F_{\pi}^2} 
  \alpha_P^{\textit{(tree)}} 
  \left( \frac{\delta F_P}{F_P^2} + 2 \frac{\delta F_\pi}{F_P^2} \right) \,,
\eeq
where the $\delta F_P$ are defined by
$F_P = \decay(1 + \delta F_P / \decay^2)$ with \cite{BB1}
\beqa \label{eq:decaycon}
\delta F_{\pi}   & = & 4 \cbeta{4}{0} (2 m_K^2 + m_\pi^2) + 4 \cbeta{5}{0} m_\pi^2
                       - \tad{\pi} - \frac{1}{2} \tad{K} \,, \no \\
\delta F_{\eta}  & = & 4 \cbeta{4}{0} (2 m_K^2 + m_\pi^2) + 4 \cbeta{5}{0} m_\eta^2
                       - \frac{3}{2} \tad{K} \,, \no \\
\delta F_{\eta'} & = & 4 (2 m_K^2 + m_\pi^2) \big(\cbeta{4}{0} + \frac{1}{3} \cbeta{5}{0}
                       - 3 \cbeta{17}{0} + \cbeta{18}{0} + \cbeta{46}{0} + 3 \cbeta{47}{0}
                       - \cbeta{53}{0} - \sqrt{\frac{3}{2}} \cbeta{52}{1} \big) \ .
\eeqa
In the case of the $\eta'$ we employed the QCD renormalization scale invariant expression
$F_{\eta'}$ which is derived from the singlet axial-vector matrix element of the $\eta'$ and
the $\cbeta{i}{0}$ are couplings of the natural parity part of the Lagrangian
at fourth chiral order, see \cite{BN, BB1} for details.

Wave-function renormalization for the external particles and
$\eta$-$\eta'$ mixing yield
\beq \label{eq:AZ}
\mathcal{A}^{\textit{(Z)}} (P \rightarrow \pi^+ \pi^- \gamma) =
  - e k_\mu \epsilon_\nu p^{+}_\alpha p^{-}_\beta
  \epsilon^{\mu \nu \alpha \beta } \dfrac{1}{4 \pi^2 F_P F_{\pi}^2}
  \beta_P^{\textit{(Z)}}
\eeq
with 
\beqa
\beta_{\eta}^{\textit{(Z)}} & = & 
  \alpha_{\eta}^{\textit{(tree)}} \Bigl(R_{8 \eta}^{(2)} + 2 R_{\pi}^{(2)} \Bigr)
  + \alpha_{\eta'}^{\textit{(tree)}} R_{0 \eta}^{(2)} \,, \no \\[1ex]
\beta_{\eta'}^{\textit{(Z)}} & = & 
  \alpha_{\eta}^{\textit{(tree)}} R_{8 \eta'}^{(2)}
  + \alpha_{\eta'}^{\textit{(tree)}} \Bigl( R_{0 \eta'}^{(2)} + 2 R_{\pi}^{(2)} \Bigr) \, ,
\eeqa
and the $R^{(2)}$ can be found in \cite{BB1}.

\subsection{Counter terms at $\mbox{\boldmath$\mathcal{O}(p^6)$}$}

The full set of contributing contact terms of unnatural parity at order $p^6$ in the
$U(3)$ framework is presented in 
App.~\ref{app:ct}. Expanding both the potentials and $U$ in the meson fields one obtains ´
(Eqs.~(\ref{eq:ctm}), (\ref{eq:ctp}))
{\arraycolsep3pt
\beqa \label{eq:unnatp6}
d^4x \mathcal{L}_{ct}^{(6)} & = &
    i\,\bar{w}_{7}^{(0)} \,\frac{8 \sqrt{2}}{f^3} 
      \trf{\{\chi \,,\phi\} (\{d\phi \,d\phi \,, dv\} + 2 d\phi \,dv \,d\phi)} \no \\
& & +\, i\,\bar{w}_{8}^{(0)} \,\frac{32 \sqrt{2}}{f^3} \trf{\chi \,\phi}
      \trf{d\phi \,d\phi \,dv} 
    -\, i\,\bar{w}_{9}^{(0)} \,\frac{16 \sqrt{6}}{f^3} \,\eta_0 
       \trf{\{\chi \,,d\phi\}[d\phi \,,dv]}
\no \\
& & -\, i\,\bar{w}_{10}^{(0)}\,\frac{16 \sqrt{6}}{f^3} \,\eta_0 
      \trf{\chi} \trf{d\phi \,d\phi \,dv} \no \\
& & -\,i \,\bar{w}_{11}^{(0)} \,\frac{16 \sqrt{2}}{f^3} 
       \trf{(\partial^\lambda d\phi \,d\phi \,\partial_\lambda \phi 
            + \partial_\lambda \phi \,d\phi \,\partial^\lambda d\phi) \,dv} \no \\
& & -\,i \,\bar{w}_{12}^{(0)} \,\frac{16 \sqrt{2}}{f^3}
       \trf{(\partial^\lambda d\phi \,\partial_\lambda \phi \,d\phi
            + d\phi \,\partial_\lambda \phi \,\partial^\lambda d\phi) \,dv} \no \\
& & -\,i \,\bar{w}_{13}^{(0)} \,\frac{16 \sqrt{2}}{f^3} \trf{d\phi}
       \trf{[\partial^\lambda d\phi \,, \partial_\lambda \phi] \,dv} 
    -\,i \,\bar{w}_{14}^{(0)} \,\frac{16 \sqrt{2}}{f^3} \trf{\partial_\lambda \phi}
       \trf{[\partial^\lambda d\phi \,, d\phi] \,dv} \ .
\eeqa}%
The first four terms contain the quark mass matrix $\mathcal{M} = \diag{(\hat{m}, \hat{m}, m_s)}$,
$\hat{m} = m_u = m_d$, which enters in the combination $\chi = 2 B \mathcal{M}$ with 
$B = - \bra{0} \bar{q} q \ket{0} / \decay^2$ being the order
parameter of the spontaneous symmetry violation, whereas
the terms $\bar{w}_{11}, \dots, \bar{w}_{14}$ arise from the counter terms with five derivatives,
since the external vector field counts as order $p$.
The resulting counter term contributions to the decay amplitudes read
\beq \label{eq:Act}
\mathcal{A}^{\textit{(ct)}} (P \rightarrow \pi^+ \pi^- \gamma) =
  - e k_\mu \epsilon_\nu p^{+}_\alpha p^{-}_\beta
  \epsilon^{\mu \nu \alpha \beta } \dfrac{1}{4 \pi^2 f^3} \ 
  \beta_P^{\textit{(ct)}}
\eeq
with 
{\arraycolsep2pt \beqa
\beta_{\eta}^{\textit{(ct)}} & = & \frac{64 \pi^2}{\sqrt{3}} \Big\{
        -4 \bar{w}_{7}^{(0)} m_{\pi}^2 + 8 \bar{w}_{8}^{(0)} (m_{K}^2 - m_{\pi}^2) \no \\
& & \qq \qq + \bar{w}_{11}^{(0)} (m_{\eta}^2 + 2 s_{+-} - 2 m_{\pi}^2)
        + \bar{w}_{12}^{(0)} (s_{+-} - 2 m_{\pi}^2) \Big\} \,, 
\no \\
\beta_{\eta'}^{\textit{(ct)}} & = & 32 \pi^2 \sqrt{\frac{2}{3}} \Big\{
        8(-\bar{w}_{7}^{(0)} + 3 \bar{w}_{9}^{(0)}) m_{\pi}^2 
        + (4 \bar{w}_{8}^{(0)} + 6 \bar{w}_{10}^{(0)}) (2 m_{K}^2 + m_{\pi}^2) \no \\
& & \qq \qq + 2 \bar{w}_{11}^{(0)} (m_{\eta'}^2 + 2 s_{+-} - 2 m_{\pi}^2)
        + 3 \bar{w}_{14}^{(0)} (m_{\eta'}^2 + s_{+-}) \no \\
& & \qq \qq + 2 (\bar{w}_{12}^{(0)} + 3 \bar{w}_{13}^{(0)}) (s_{+-} - 2 m_{\pi}^2) \Big\} \ .
\eeqa}%
The $\bar{w}_i^{(0)}$ are understood to be the finite parts of the coupling constants in the 
Lagrangian of sixth chiral order. Within $SU(3)$ ChPT the absorption of divergences from 
one-loop diagrams into the LECs of the $\mathcal{O}(p^6)$ Lagrangian has been discussed in 
\cite{BBC}.
For the numerical analysis
it is convenient to decompose the $\beta_{P}^{\textit{(ct)}}$ as follows,
\beq
\beta_{\eta}^{\textit{(ct)}} = \frac{64 \pi^2}{\sqrt{3}} \bigl( 
  \bar{w}_{\eta}^{(m)} + \bar{w}_{\eta}^{(s)} s_{+-} \bigr)
\eeq
with
\beqa \label{eq:ctcombeta}
\bar{w}_{\eta}^{(m)} & = & (-4 \bar{w}_{7}^{(0)} - 2 \bar{w}_{11}^{(0)} - 2 \bar{w}_{12}^{(0)}) m_{\pi}^2 
  + 8 \bar{w}_{8}^{(0)} (m_{K}^2 - m_{\pi}^2) + \bar{w}_{11}^{(0)} m_{\eta}^2 \,, \no \\
\bar{w}_{\eta}^{(s)} & = & 2 \bar{w}_{11}^{(0)} + \bar{w}_{12}^{(0)} \ ,
\eeqa
and
\beq
\beta_{\eta'}^{\textit{(ct)}} = 32 \pi^2 \sqrt{\frac{2}{3}} \bigl( 
  \bar{w}_{\eta'}^{(0)} m_{\eta'}^2  + \bar{w}_{\eta'}^{(m)} 
  + \bar{w}_{\eta'}^{(s)} s_{+-} \bigr)
\eeq
with
\beqa \label{eq:ctcombetap}
\bar{w}_{\eta'}^{(0)} & = &  2 \bar{w}_{11}^{(0)} + 3 \bar{w}_{14}^{(0)} \no \\ 
\bar{w}_{\eta'}^{(m)} & = & -4(2 \bar{w}_{7}^{(0)} - 6 \bar{w}_{9}^{(0)} 
  + \bar{w}_{11}^{(0)} + \bar{w}_{12}^{(0)} + 3 \bar{w}_{13}^{(0)}) m_{\pi}^2 
+ (4 \bar{w}_{8}^{(0)} + 6 \bar{w}_{10}^{(0)}) (2 m_{K}^2 + m_{\pi}^2)  \,, \no \\
\bar{w}_{\eta'}^{(s)} & = & 
  4 \bar{w}_{11}^{(0)} + 2 \bar{w}_{12}^{(0)} + 6 \bar{w}_{13}^{(0)} + 3 \bar{w}_{14}^{(0)} \ .
\eeqa
The $\bar{w}_{\eta'}^{(0)}$ piece in $\beta_{\eta'}^{\textit{(ct)}}$ 
can be absorbed into $w_{3}^{(1)}$
\beq
w_{3}^{(1)} = w_{3}^{(1)r} + \frac{1}{16 \pi^2} \bar{w}_{\eta'}^{(0)} m_{\eta'}^2 - \ldots \ ,
\eeq
where the ellipsis denotes the renormalization of the one-loop graphs given
in Eq.~(\ref{eq:ren}) and in the following we employ the redefined value
for  $\beta_{\eta'}^{\textit{(ct)}}$ after renormalization,
$\beta_{\eta'}^{\textit{(ct)}} =  32 \pi^2 \sqrt{\frac{2}{3}} \bigl( 
\bar{w}_{\eta'}^{(m)} + \bar{w}_{\eta'}^{(s)} s_{+-} \bigr)$,
without changing the notation.


\subsection{Full one-loop result}

Summing up the contributions up to $\mathcal{O}(p^6)$ we arrive at 
\beq
\mathcal{A}^{\textit{(1-loop)}} = \mathcal{A}^{\textit{(tree)}} 
  + \mathcal{A}^{\textit{(tad)}} + \mathcal{A}^{\textit{(a)}}
  + \mathcal{A}^{\textit{(b)}} + \mathcal{A}^{\textit{(c)}}
  + \mathcal{A}^{\textit{(f)}} + \mathcal{A}^{\textit{(Z)}}
  + \mathcal{A}^{\textit{(ct)}} \ .
\eeq
The amplitude has the form
\beq  \label{eq:ANLO}
\mathcal{A}^{\textit{(1-loop)}} (P \rightarrow \pi^+ \pi^- \gamma) =
  - e k_\mu \epsilon_\nu p^{+}_\alpha p^{-}_\beta
  \epsilon^{\mu \nu \alpha \beta } \dfrac{1}{4 \pi^2 F_P F_{\pi}^2} \ 
  \beta_P^{\textit{(1-loop)}}
\eeq
with 
\beqa
\beta_P^{\textit{(1-loop)}} & = & 
  \alpha_P^{\textit{(tree)}} 
  \left(1 + \frac{\delta F_P}{F_P^2} + 2 \frac{\delta F_\pi}{F_P^2} \right)
  + \beta_P^{\textit{(tad)},\pi} \, \frac{\Delta_\pi}{F_P^2}
  + (\beta_P^{\textit{(tad)},K} + \beta_P^{\textit{(b)}} ) 
    \frac{\Delta_K}{F_P^2} \no \\ 
& & + \frac{\beta_P^{\textit{(a)},\pi}}{F_P^2} \,I_1(m_{\pi}^2; s_{+-})
  + \frac{\beta_P^{\textit{(a)},K}}{F_P^2}  \,I_1(m_{K}^2; s_{+-}) \no \\
& & + \frac{\beta_P^{\textit{(c)}}}{F_P^2} \,
  \big(\hat{I}_2(m_{\pi}^2, m_{\eta'}^2; s_{+ \gamma}) 
      + \hat{I}_2(m_{\pi}^2, m_{\eta'}^2; s_{- \gamma}) \big)
  + \beta_P^{(Z)} + \beta_P^{(ct)} \, ,
\eeqa
where we have replaced the 
decay constant in the chiral limit, $f$, by the decay constants $F_P$ 
($P = \eta, \eta'$) and $F_\pi$, Eq.~(\ref{eq:decaycon}),
in such a way that the amplitude $\mathcal{A}^{\textit{(1-loop)}}$ is accompanied by a factor 
$1/(F_P F_{\pi}^2)$.  
Note that in $\alpha_P^{\textit{(tree)}}$ the coefficient $w_{3}^{(1)}$ has been replaced by
$w_{3}^{(1)r}$, in order to account for the renormalization of the $p^4$ contributions from
loop graphs and counter terms.
For convenience we also show the explicit
form of the $\beta_P^{\textit{(1-loop)}}$
{\arraycolsep2pt \beqa \label{eq:NLOcoeff}
\beta_{\eta}^{\textit{(1-loop)}} & = & 
  \frac{1}{\sqrt{3}} \Bigg\{ 1 + \frac{1}{F_{\eta}^2} \Bigg[4 \sqrt{\frac{2}{3}} 
  \left(\sqrt{\frac{2}{3}} - 16 \pi^2 w_{3}^{(1)r} \right) (m_{K}^2 - m_{\pi}^2) 
  \frac{\cvtwid{2}{1}}{v_{0}^{(2)}}  \no \\
& & \qq - 3 \ \Delta_\pi + I_1(m_{\pi}^2; s_{+-}) + 2 I_1(m_{K}^2; s_{+-}) \Bigg] 
    + 64 \pi^2 \bigl( \bar{w}_{\eta}^{(m)} + \bar{w}_{\eta}^{(s)} s_{+-} \bigr) \Bigg\} \,, 
\no \\
\beta_{\eta'}^{\textit{(1-loop)}} & = &
  \left(\sqrt{\frac{2}{3}} - 16 \pi^2 w_{3}^{(1)r} \right) \Bigg\{ 1 + 
  \frac{1}{F_{\eta'}^2} \Bigg[ 4 (2 m_{K}^2 + m_{\pi}^2)
  \Bigl(\cbeta{46}{0} +3\cbeta{47}{0} -\cbeta{53}{0} -\sqrt{\frac{3}{2}} \cbeta{52}{1}\Bigr)
  \no \\
& & \qq - \ 3 \Delta_\pi - \frac{3}{2} \Delta_K 
  + I_1(m_{\pi}^2; s_{+-}) + \frac{1}{2} I_1(m_{K}^2; s_{+-}) \no \\
& & \qq - \ 4 v_{1}^{(2)} \bigl(\hat{I}_2(m_{\pi}^2, m_{\eta'}^2; s_{+ \gamma}) 
  + \hat{I}_2(m_{\pi}^2, m_{\eta'}^2; s_{- \gamma}) \bigr) \Bigg] \Bigg\} \no \\
& & + \frac{4}{3} \sqrt{\frac{2}{3}} (m_{K}^2 - m_{\pi}^2) \left(
  4 \frac{\beta_{5, 18}}{F_{\eta'}^2} - \frac{\cvtwid{2}{1}}{v_{0}^{(2)}} \right) 
  + 32 \pi^2 \sqrt{\frac{2}{3}} 
  \bigl( \bar{w}_{\eta'}^{(m)} + \bar{w}_{\eta'}^{(s)} s_{+-} \bigl) \ ,
\eeqa}%
where we have used the abbreviations
\beqa
\cvtwid{2}{1} & = & \sfrac{1}{4} f^2 - \sfrac{1}{2}\sqrt{6}\coeffv{3}{1} \,, \no \\
\beta_{5,18}  & = & \cbeta{5}{0}+\sfrac{3}{2}\cbeta{18}{0} \ .
\eeqa
This generalizes the result from \cite{BBC} to the $U(3)$ framework
with the $\eta'$ as an explicit degree of freedom.

\section{Generation of resonances}  \label{sec:CC}

\begin{figure}
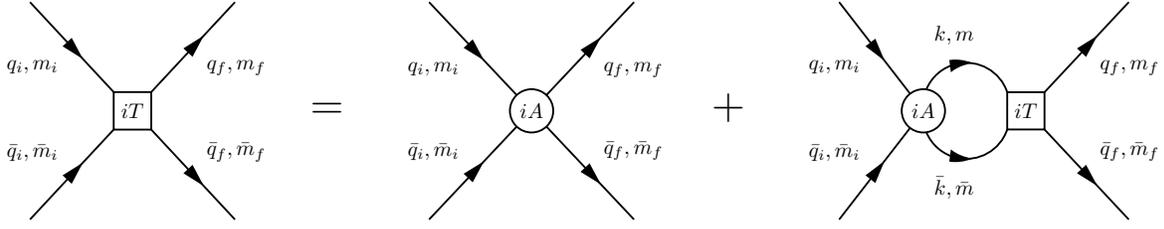

\centering
\begin{tabular}{ccccc}
\parbox{0.19\textwidth}{\includegraphics[scale=0.8]{feynps.5}} & \quad {\Large =} & \quad
\parbox{0.19\textwidth}{\includegraphics[scale=0.8]{feynps.6}} & \quad {\Large +} & \quad
\parbox{0.27\textwidth}{\includegraphics[scale=0.8]{feynps.9}}
\end{tabular}
\caption{Diagrammatic Bethe-Salpeter equation for meson-meson rescattering.} 
\label{fig:BSE}
\end{figure}

Resonances, in particular the $\rho(770)$, 
play an important role in the decays of $\eta$ and $\eta'$ into $\pi^+ \pi^- \gamma$.
We include them in the same way as for the two-photon decays in \cite{BN}, 
{\it i.e.}\ by employing a coupled channel Bethe-Salpeter equation (BSE) which 
satisfies unitarity constraints and generates resonances dynamically by an infinite string of  
meson-meson rescattering processes. The scattering potential $A$ is calculated from the contact 
interactions of the effective Lagrangian up to fourth chiral order and iterated in the
coupled channel Bethe-Salpeter equation 
\beq
T = [\mathds{1} + A \cdot \tilde{G}]^{-1} \; A
\eeq
with a modified scalar loop integral
\beq
\tilde{G}_{m \bar{m}} = G_{m \bar{m}} (\mu) + a_{m \bar{m}} (\mu) \,,
\eeq
where $G$ has been defined in Eq.~(\ref{eq:intG}) and $a$ is a real constant chosen in such 
a way that the $\mu$ dependent pieces of $G$ and $a$ compensate. 
For a diagrammatic illustration of the BSE, see Fig.~\ref{fig:BSE}.
The resulting $T$ matrix describes accurately the experimental phase shifts in both the $s$ and 
$p$ wave channels \cite{BN, BB4}. 

\begin{figure}
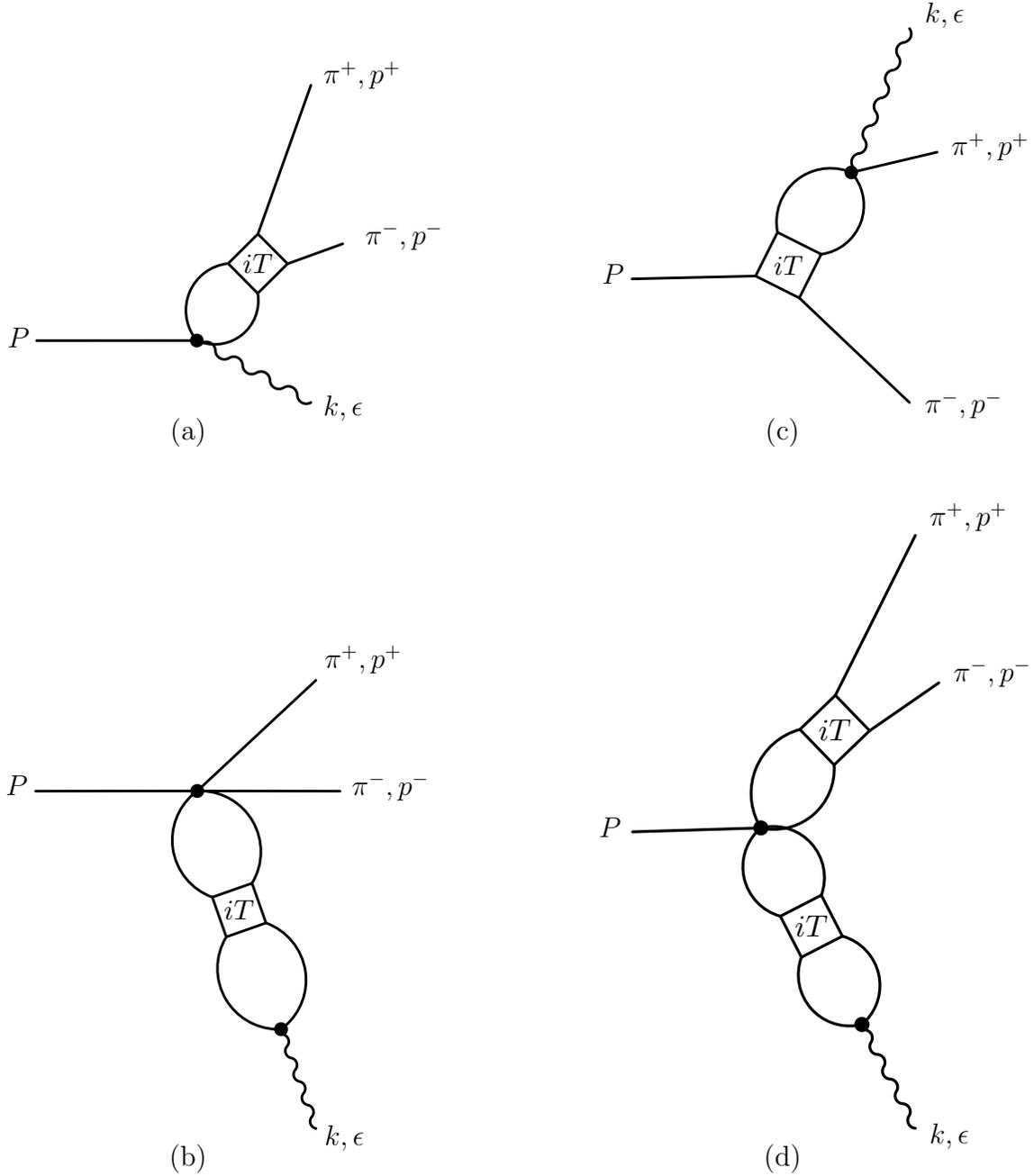

\begin{center}
\begin{minipage}[b]{0.3\textwidth}
\centering
\begin{overpic}[width=0.9\textwidth]{feynps.19}
\put(-8,18){\scalebox{1.0}{$P$}}
\put(103,50){\scalebox{1.0}{$\pi^-, p^-$}}
\put(90,100){\scalebox{1.0}{$\pi^+, p^+$}}
\put(90,-3){\scalebox{1.0}{$k, \epsilon$}}
\end{overpic} \\
(a) 
\end{minipage}
\hspace{0.2\textwidth}
\begin{minipage}[b]{0.3\textwidth}
\centering
\begin{overpic}[width=0.9\textwidth]{feynps.21}
\put(-7,32){\scalebox{1.0}{$P$}}
\put(78,101){\scalebox{1.0}{$k, \epsilon$}}
\put(85,66){\scalebox{1.0}{$\pi^+, p^+$}}
\put(78,-2){\scalebox{1.0}{$\pi^-, p^-$}}
\end{overpic} \\
(c)
\end{minipage}
\end{center}
\hspace{2cm}
\begin{center}
\begin{minipage}[b]{0.3\textwidth}
\centering
\begin{overpic}[width=0.9\textwidth]{feynps.20}
\put(-6,74){\scalebox{1.0}{$P$}}
\put(64,-3){\scalebox{1.0}{$k, \epsilon$}}
\put(64,102){\scalebox{1.0}{$\pi^+, p^+$}}
\put(70,74){\scalebox{1.0}{$\pi^-, p^-$}}
\end{overpic} \\
(b)
\end{minipage}
\hspace{0.2\textwidth}
\begin{minipage}[b]{0.3\textwidth}
\centering
\begin{overpic}[width=0.9\textwidth]{feynps.22}
\put(-5,49){\scalebox{1.0}{$P$}}
\put(50,-2){\scalebox{1.0}{$k, \epsilon$}}
\put(50,101){\scalebox{1.0}{$\pi^+, p^+$}}
\put(54,75){\scalebox{1.0}{$\pi^-, p^-$}}
\end{overpic} \\
(d)
\end{minipage}
\end{center}
\caption{Set of meson-meson rescattering processes in the decay $P \to \pi^+ \pi^- \gamma$ 
         included in this approach. The crossed diagram of (c) is not shown.}
\label{fig:CC}
\end{figure}

In order to implement the non-perturbative summation of loop graphs 
covered by the BSE in the decay processes
under consideration, we treat the BSE $T$ matrix as an effective vertex for meson-meson scattering. 
The pertinent diagrams are shown in Fig.~\ref{fig:CC}
and an analysis of the remaining loop integrations confirms that---as in the case of the 
two-photon decays---only the $p$ wave part of $T$ contributes.

There is, however, one difference to the two-photon decays. Whereas for the process
$P \to \gamma^{(*)} \gamma^{(*)}$ the contributions of the coupled channels were at least 
of two-loop order, the diagrams (a) and (c) in Fig.~\ref{fig:CC} incorporate also one-loop
contributions, since $T = A$ at lowest order. In order to avoid double counting, we omit 
those one-loop contributions from the last section which are already covered by the coupled 
channel calculation. Since the coupled channel diagrams (a) and (c) exactly reproduce the one-loop 
results at lowest order, the next-to-leading order calculation is still complete.

%
The amplitude of Fig.~\ref{fig:CC}a is given by
\beq  \label{eq:ACCa}
\mathcal{A}^{\textit{(CCa)}}(P \to \gamma^{(*)} \gamma^{(*)}) =  
 - e k_\mu \epsilon_\nu p^{+}_\alpha p^{-}_\beta \epsilon^{\mu \nu \alpha \beta }
\frac{1}{4 \pi^2 f^3}
{\sum_{a}}' \gamma_{P}^{\textit{(CCa)},a} 
\ \tilde{I}_1(m_{a}^2; s_{+-}) \ \hat{T}_{p}^{(a \to \pi^\pm)}(s_{+-}) 
\eeq
with 
{\arraycolsep2pt \beqa
\gamma_{\eta}^{\textit{(CCa)},\pi^{\pm}} 
& = & \gamma_{\eta}^{\textit{(CCa)},K^{\pm}}
= \dfrac{1}{6} \left[\sqrt{3} + \dfrac{4\sqrt{2}}{3}(m_{K}^2 - m_{\pi}^2)
\dfrac{\cvtwid{2}{1}}{\coeffv{0}{2}} \bigl(\sqrt{6} - 48 \pi^2 w_{3}^{(1)r} \bigr) \right] , \no \\
\gamma_{\eta}^{\textit{(CCa)},K^0 \bar{K}^0} & = & - \dfrac{\sqrt{3}}{2} \,, \no \\[2ex]
\gamma_{\eta'}^{\textit{(CCa)},\pi^{\pm}} 
& = & \gamma_{\eta'}^{\textit{(CCa)},K^{\pm}}
\no \\
& = & \dfrac{1}{6} \left[\sqrt{6} - 48 \pi^2 w_{3}^{(1)r} + \dfrac{4\sqrt{6}}{3}
(m_{K}^2 - m_{\pi}^2) \left( 4 \dfrac{\beta_{5,18}}{F_{\eta'}^2} 
  - \dfrac{\cvtwid{2}{1}}{\coeffv{0}{2}} \right) \right] , \no \\
\gamma_{\eta'}^{\textit{(CCa)},K^0 \bar{K}^0} 
& = & - 2 \sqrt{\dfrac{2}{3}} (m_{K}^2 - m_{\pi}^2) \left(
      4 \dfrac{\beta_{5,18}}{F_{\eta'}^2} - \dfrac{\cvtwid{2}{1}}{\coeffv{0}{2}} \right) \ .
\eeqa}%
The symbol ${\sum}'$ in Eq.~(\ref{eq:ACCa}) denotes summation over 
the meson pairs $\pi^+ \pi^-$, $K^+ K^-$ and $K^0 \bar{K}^0$ and 
$\hat{T}_{p}^{(a \to b)}$ is the $p$ wave part of the BSE $T$ matrix for scattering of a 
meson pair $a$ into a meson pair $b$ as defined in \cite{BN}. 
The loop integral $\tilde{I}_1$ is given by
\begin{equation} \label{eq:intI1t}
\tilde{I}_1(m_\phi^2;p^2) = I_1(m_\phi^2;p^2) + C_\phi \ p^2 
\end{equation}
with $I_1$ defined in Eq.~(\ref{eq:intI1}) and
in order to be in better agreement with experiment
we make use of the freedom to take arbitrary values for the analytic pieces
of the integrals which corresponds to a specific choice
of counter term contributions.

The diagram in Fig.~\ref{fig:CC}b yields the amplitude
\begin{multline}  \label{eq:ACCb}
\mathcal{A}^{\textit{(CCb)}}(P \to \gamma^{(*)} \gamma^{(*)}) =  
 - e k_\mu \epsilon_\nu p^{+}_\alpha p^{-}_\beta \epsilon^{\mu \nu \alpha \beta }
   \frac{1}{4 \pi^2 f^5} {\sum_{a}}' \gamma_{P}^{\textit{(CCb)},a} 
   \Delta_K \\[1ex]
\times \left[\hat{T}_{p}^{(a \to \pi^\pm)}(0) \ \Delta_\pi 
   + \hat{T}_{p}^{(a \to K^\pm)}(0) \ \Delta_K  \right] \,,
\end{multline}
where the coefficients $\gamma_{P}^{\textit{(CCb)},a}$ are given by
{\arraycolsep2pt \beqa
\gamma_{\eta}^{\textit{(CCb)},\pi^{\pm}} 
& = & \gamma_{\eta'}^{\textit{(CCb)},\pi^{\pm}} = 0 \,, \qq
\gamma_{\eta}^{\textit{(CCb)},K^{\pm}}
= -\gamma_{\eta}^{\textit{(CCb)},K^0 \bar{K}^0}
=  \dfrac{\sqrt{3}}{2} \,, \no \\[1ex]
\gamma_{\eta'}^{\textit{(CCb)},K^{\pm}}
& = & -\gamma_{\eta'}^{\textit{(CCb)},K^0 \bar{K}^0} 
= 2 \sqrt{\dfrac{2}{3}} (m_{K}^2 - m_{\pi}^2) \left(
      4 \dfrac{\beta_{5,18}}{F_{\eta'}^2} - \dfrac{\cvtwid{2}{1}}{\coeffv{0}{2}} \right) \ .
\eeqa}%
The diagram in Fig.~\ref{fig:CC}c involves the $p$ wave scattering matrix for the two coupled 
channels $\ket{\pi \eta}$ and $\ket{\pi \eta'}$, so that the loop integral contains a pion
and either an $\eta$ or an $\eta'$.\footnote{The $(\pi, \eta)$ loop stems from the 
$\mathcal{O}(p^4)$ part of the scattering potential $A$ and did thus not appear in the 
next-to-leading order calculation, where only the four-meson vertex of second chiral order 
was considered.}  
The amplitude from 
Fig.~\ref{fig:CC}c and the crossed diagram reads
\begin{multline}  \label{eq:ACCc}
\mathcal{A}^{\textit{(CCc)}}(P \to \gamma^{(*)} \gamma^{(*)}) =  
 - e k_\mu \epsilon_\nu p^{+}_\alpha p^{-}_\beta \epsilon^{\mu \nu \alpha \beta }
  \frac{1}{4 \pi^2 f^3} \\
\times \frac{1}{2} \Bigg\{ \frac{1}{\sqrt{3}} \big[
  I_2(m_{\pi}^2, m_{\eta}^2; s_{+ \gamma}) 
  \ \hat{T}_{p}^{(P \pi^+ \to \eta \pi^+)}(s_{+ \gamma}) 
+ I_2(m_{\pi}^2, m_{\eta}^2; s_{- \gamma})
  \ \hat{T}_{p}^{(P \pi^- \to \eta \pi^-)}(s_{- \gamma}) \big] \\
\shoveleft \qq \qq \q + \Bigl(\sqrt{\frac{2}{3}} - 16 \pi^2 w_{3}^{(1)r} \Bigr) \big[
  \hat{I}_2(m_{\pi}^2, m_{\eta'}^2; s_{+ \gamma})
  \ \hat{T}_{p}^{(P \pi^+ \to \eta' \pi^+)}(s_{+ \gamma}) \\
+ \hat{I}_2(m_{\pi}^2, m_{\eta'}^2; s_{- \gamma})
  \ \hat{T}_{p}^{(P \pi^- \to \eta' \pi^-)}(s_{- \gamma}) \big] \Bigg\} \,,
\end{multline}
and the integrals $I_2$ and $\hat{I}_2$ have been defined in Eq.~(\ref{eq:intI2})
and Eq.~(\ref{eq:intI2p}), respectively.

We also consider the diagram with two coupled channels depicted in 
Fig.~\ref{fig:CC}d. The pertinent amplitude reads
\begin{multline} \label{eq:A2CC}
\mathcal{A}^{\textit{(2\,CC)}}(P \to \gamma^{(*)} \gamma^{(*)}) =
  - e k_\mu \epsilon_\nu p^{+}_\alpha p^{-}_\beta \epsilon^{\mu \nu \alpha \beta }
\frac{1}{4 \pi^2 f^5}
{\sum_{a,b}}' \gamma_{P}^{\textit{(2\,CC)},a,b} \\[1ex]
\times \,\tilde{I}_1(m_{a}^2; s_{+-}) \ \hat{T}_{p}^{(a \to \pi^\pm)}(s_{+-}) \ \Delta_b 
  \big[ \hat{T}_{p}^{(b \to  \pi^\pm)}(0) \ \Delta_\pi
  + \hat{T}_{p}^{(b \to K^\pm)}(0) \ \Delta_K \big] \ .
\end{multline}
with coefficients $\gamma_{P}^{\textit{(2\,CC)},a,b}$ symmetric under 
$a \leftrightarrow b$
{\arraycolsep2pt \beqa
\gamma_{\eta}^{\textit{(2\,CC)},\pi^{\pm},K^{\pm}}
& = & - \gamma_{\eta}^{\textit{(2\,CC)},\pi^{\pm},K^0 \bar{K}^0}
= - \dfrac{1}{2} \gamma_{\eta}^{\textit{(2\,CC)},K^{\pm},K^0 \bar{K}^0}
= \dfrac{\sqrt{3}}{4} \,, \no \\[2ex]
\gamma_{\eta'}^{\textit{(2\,CC)},\pi^{\pm},K^{\pm}}
& = & - \gamma_{\eta'}^{\textit{(2\,CC)},\pi^{\pm},K^0 \bar{K}^0}
= - \dfrac{1}{2} \gamma_{\eta'}^{\textit{(2\,CC)},K^{\pm},K^0 \bar{K}^0} \no \\[1ex]
& = & \ \sqrt{\dfrac{2}{3}} (m_{K}^2 - m_{\pi}^2) \left(
        4 \dfrac{\beta_{5,18}}{F_{\eta'}^2} - \dfrac{\cvtwid{2}{1}}{\coeffv{0}{2}} \right) 
\eeqa}%
and zero otherwise.

\section{Numerical results} \label{sec:num}

The spectra of the decays $\eta, \eta' \to \pi^+ \pi^- \gamma$ have been measured with high
statistics \cite{Gor, Lay, GAMS, CB} showing that the $\eta'$ decay is clearly dominated by 
the $\rho$ resonance. Therefore, the next-to-leading order calculation of Sec.~\ref{sec:1loop} 
is insufficient to describe the $\eta'$ decay and we will discuss the one-loop results only for the 
$\eta$ decay. As a first estimate we use a set of parameters which is consistent with previous one-loop 
calculations in $U(3)$ ChPT \cite{BN, BB1, BB2}
\beqa
F_\pi & = & 92.4 \MeV  \,, \qq F_{\eta} =  1.3 F_\pi  \,, \qq F_{\eta'} = 1.1 F_\pi \,  , \no \\
\cbeta{5}{0} & = & 1.4 \cdot 10^{-3} \,, \qq \cvtwid{2}{1} = 1.2 F_{\pi}^2 / 4 ,
\eeqa
while neglecting all other LECs including the coupling constants of the unnatural parity Lagrangian 
of sixth chiral order. 
For the regularization scale $\mu$ we use 1 GeV.
The decay width turns out to be $\Gamma_{\eta} = 14.48$~eV 
which is by a factor of four smaller than the experimental value quoted by the Particle 
Data Group \cite{pdg}
\beq
\Gamma_{\eta} = (56.1 \pm 5.4) \eV \ .
\eeq
A mixing parameter of $\cvtwid{2}{1} = 0.5 F_{\pi}^2 / 4$ corresponds to an even smaller decay 
width of $\Gamma_{\eta} = 8.59$~eV, whereas a value of $\cvtwid{2}{1} = 1.5 F_{\pi}^2 / 4$ 
results in $\Gamma_{\eta} = 17.47$~eV. 
Hence, for realistic values of the $\eta$-$\eta'$
mixing parameter $\cvtwid{2}{1}$, it is not possible to match the 
experimental decay width without including the contact terms of $\mathcal{O}(p^6)$. 

However, both the photon spectrum and the partial width of the decay
$\eta \to \pi^+ \pi^- \gamma$ can be reproduced at next-to-leading order, if $\mathcal{O}(p^6)$
counter terms are included. But the choice of parameters is not unique, 
hence it is not possible to constrain the mixing parameter
$\cvtwid{2}{1}$ from a fit to the experimental data on the $\eta$ decay,
since variations in $\cvtwid{2}{1}$ can be compensated by tuning the counter terms
of unnatural parity; larger positive values 
of $\cvtwid{2}{1}$, {\it e.g.}, require smaller counter term contributions.
For consistency with previous one-loop calculations in $U(3)$ ChPT \cite{BN, BB1, BB2} we prefer 
to use a mixing parameter of $\cvtwid{2}{1} = 1.2 F_{\pi}^2 / 4$.
At the one-loop level the constants $w_{3}^{(1)r}$ and $\bar{w}_{\eta}^{(m)}$ 
appear in a linear combination and thus cannot 
be fixed separately. In order to work with a minimum set of parameters, we set 
in the one-loop calculation $w_{3}^{(1)r}$ to zero
and choose the combinations of counter terms Eq.~(\ref{eq:ctcombeta}) to be
$\bar{w}_{\eta}^{(m)}  = -0.31 \cdot 10^{-3}$ and 
$\bar{w}_{\eta}^{(s)}  = 12.9  \cdot 10^{-3} \GeV^{-2}$.

\begin{figure}
\centering
\begin{minipage}{0.45\textwidth}
\includegraphics[width=1.0\textwidth,clip]{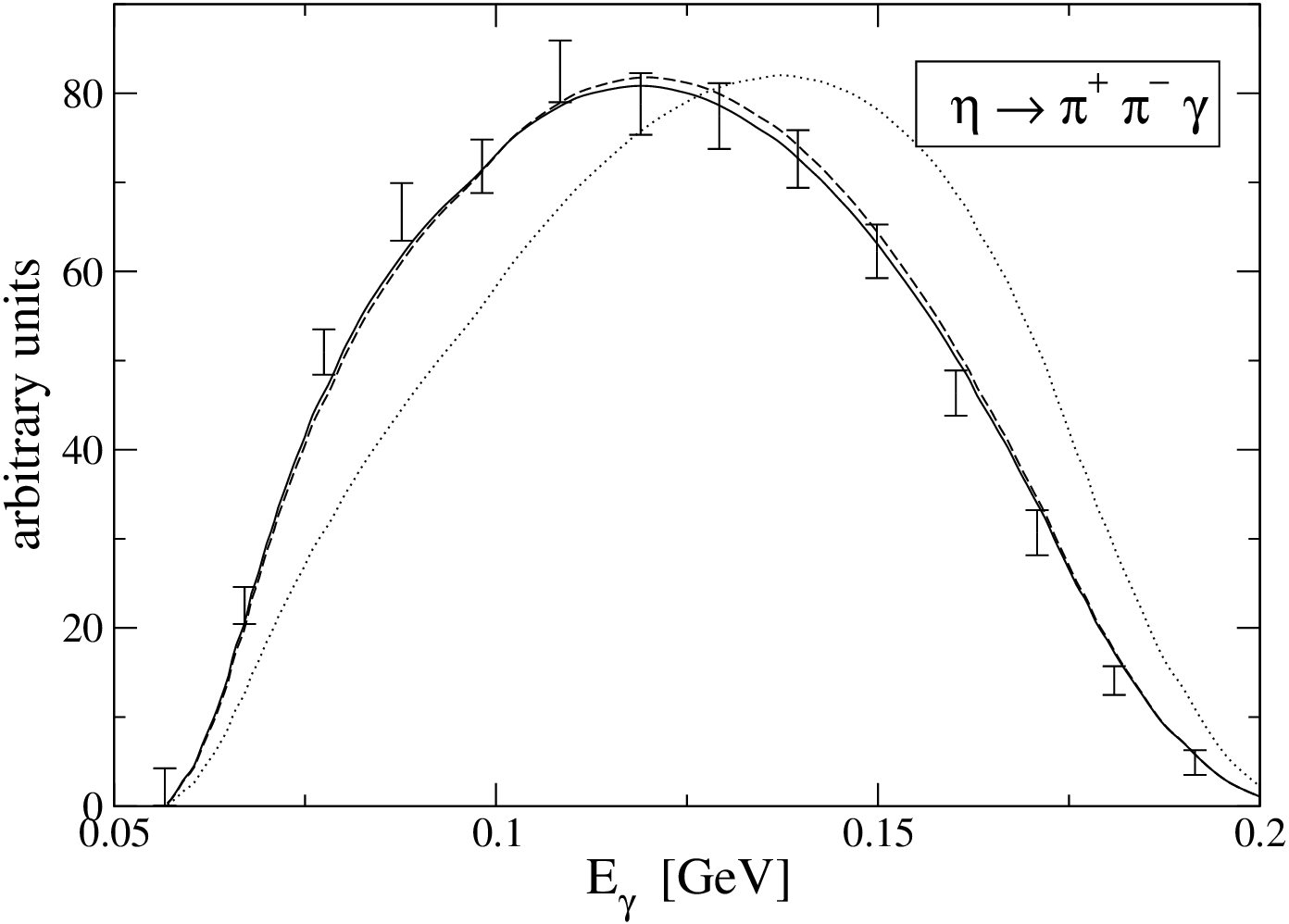}
\end{minipage}
\caption{Photon spectrum resulting from the next-to-leading order calculation 
         with $p^6$ counter terms (dashed) and from the
	 full calculation including the coupled channels (solid). 
	 The dotted line corresponds to the simplest gauge invariant amplitude
	 $\mathcal{A} \propto k_\mu \epsilon_\nu p^{+}_\alpha p^{-}_\beta 
	 \epsilon^{\mu \nu \alpha \beta }$.
         The data are taken from \cite{Gor}.}
\label{fig:PSeta}
\end{figure}

The one-loop result is compared with experimental data from \cite{Gor} in 
Fig.~\ref{fig:PSeta} and shows good agreement after accounting for the detection efficiencies.
Good agreement with the similar experiment \cite{Lay} is also achieved. 
It should be mentioned that the counter term contribution $\bar{w}_{\eta}^{(s)}$
yields an important contribution to the decay amplitude and dominates the one-loop
corrections.
This fact is a reflection of the tail of the $\rho$ resonance in the 
perturbative expansion \cite{BBC}.

\begin{figure}
\centering
\begin{minipage}{0.45\textwidth}
\begin{overpic}[width=1.0\textwidth,clip]{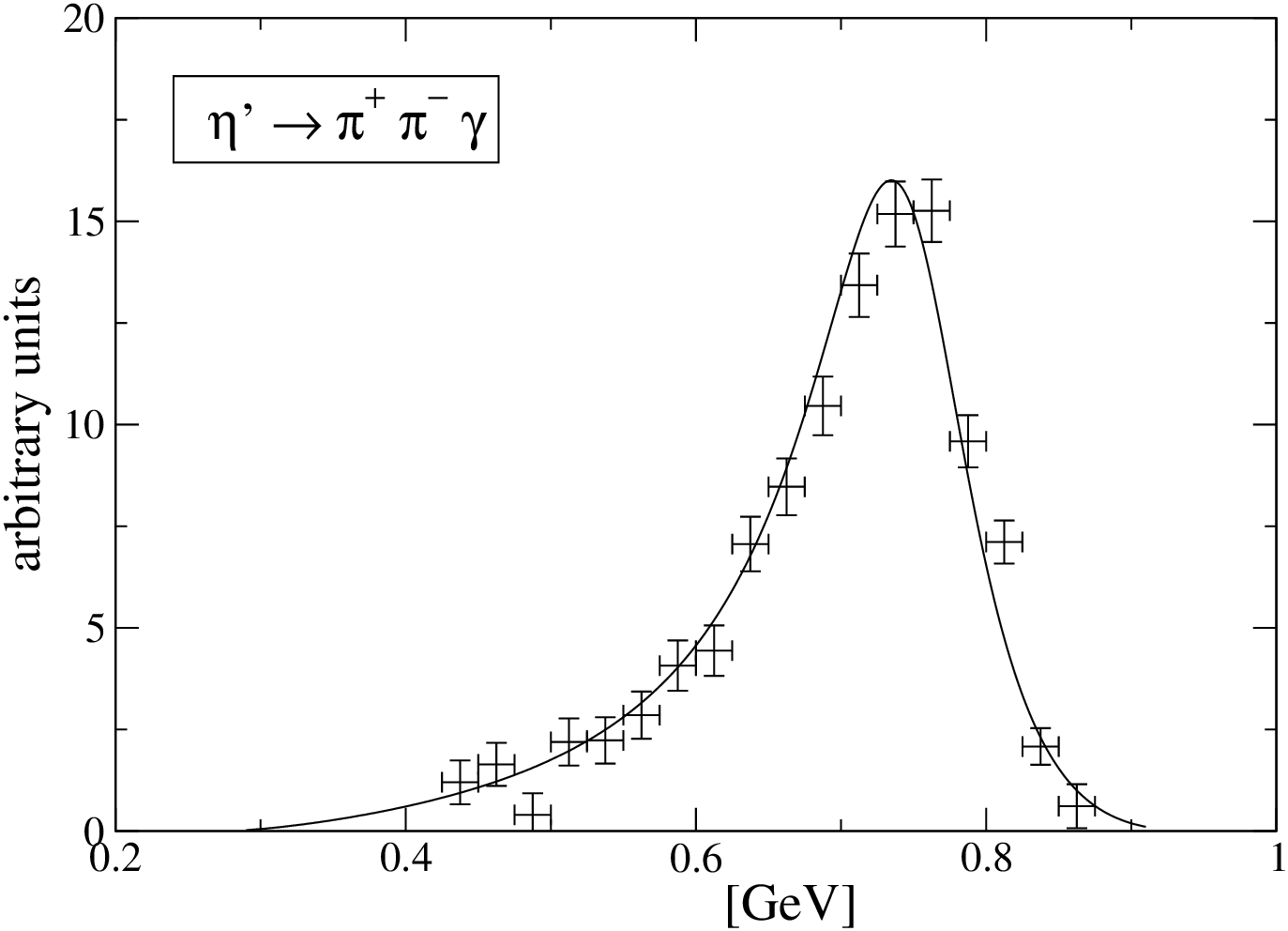}
\put(42,0.5){\scalebox{0.7}{$\sqrt{s_{+-}}$}}
\end{overpic}
\end{minipage}
\hspace{0.05\textwidth}
\begin{minipage}{0.45\textwidth}
\begin{overpic}[width=1.0\textwidth,clip]{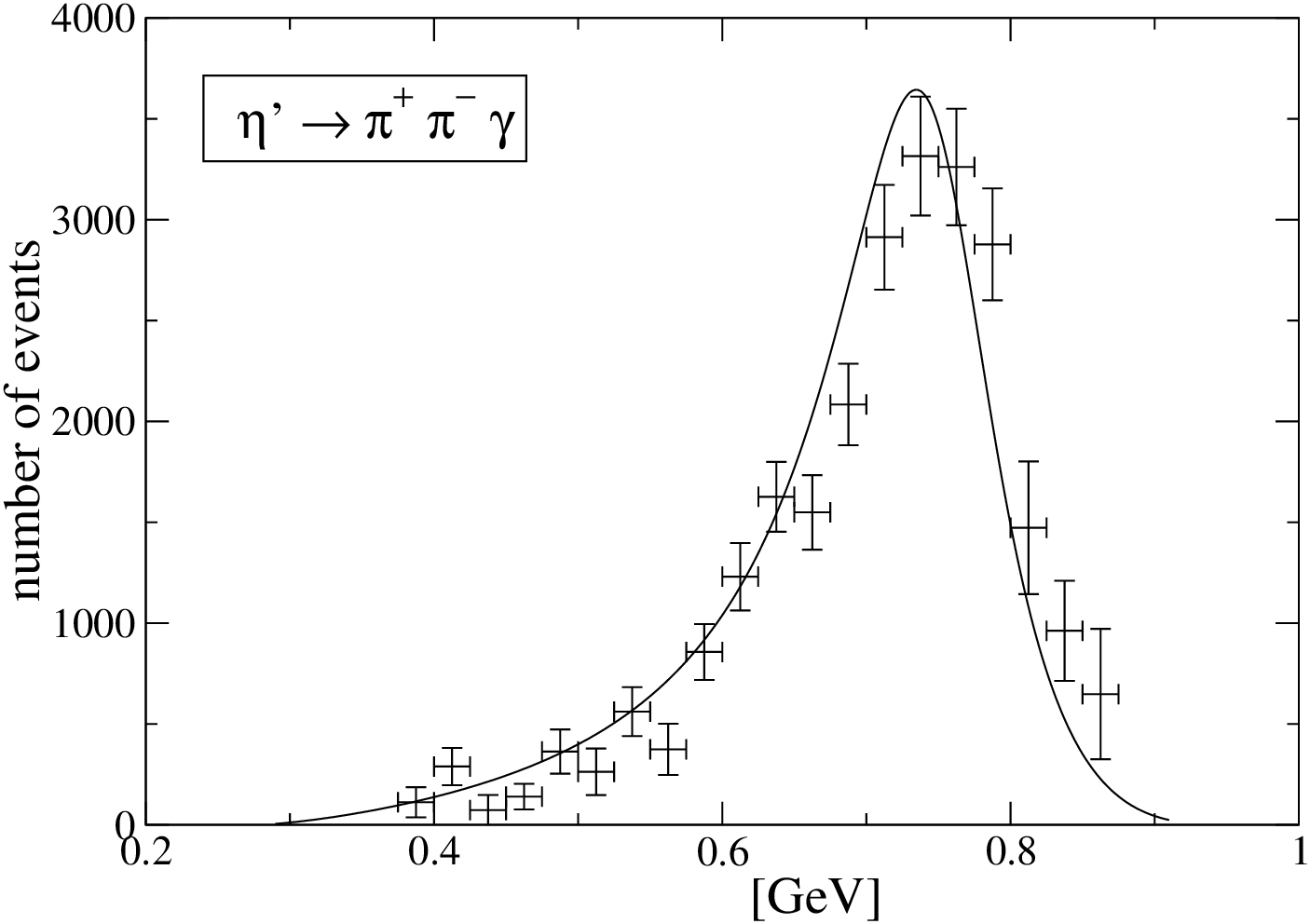}
\put(42,0.5){\scalebox{0.7}{$\sqrt{s_{+-}}$}}
\end{overpic}
\end{minipage}
\caption{Invariant mass spectrum of the $\pi^+ \pi^-$ system 
         resulting from the full calculation.
	 The curves are normalized to the integral of the experimental
         distribution.
         Data: left diagram \cite{CB}, right diagram \cite{GAMS}.}
\label{fig:IMetapbest}
\end{figure}

We now turn to the discussion  of the coupled channel calculation. The 
computation includes the tree level graphs and all next-to-leading order corrections. 
In order to avoid double counting, 
the one-loop contributions which are already covered by the coupled channel
diagrams in Figs.\ref{fig:CC}a,~c have been subtracted.
The results are plotted in Figs.~\ref{fig:PSeta},
\ref{fig:IMetapbest} and compared with the experimental data. 
The agreement with the data is very good and shows
that the important degrees of freedom are incorporated in our model. 
For consistency with previous calculations in this framework
\cite{BN, BB3}, we have set the mixing 
parameter $\cvtwid{2}{1}$ and the renormalized coupling constant $w_{3}^{(1)r}$ 
from the unnatural parity $\mathcal{O}(p^4)$ Lagrangian to zero. 
We furthermore neglect the unknown LEC $v_{1}^{(2)}$,
in order to work with a minimal set of free parameters. 
For the fit of our results to the central experimental values of the partial decay widths, 
$\Gamma_{\eta} = 56.1$~eV and $\Gamma_{\eta'} = 59.6$~keV \cite{pdg}, respectively, we employ
for the subtraction constant $C_\pi$ in the  outer loop integral with pions 
(Eq.~(\ref{eq:intI1t})) the value
$C_\pi (\mu = 1\GeV) = -1.42 \cdot 10^{-2}$ 
which is comparable in size with the one used in \cite{BN} 
($C_\pi = -1/(6 \pi^2) \approx -1.69 \cdot 10^{-2}$), while setting those
for the other loops to zero. 
Since different sets of counter term contributions are summarized in $C_\pi$, 
the values of these subtraction constants may differ for 
the decays into $2 \pi \gamma$ or two photons.
The counter term contributions from the Lagrangian 
of sixth chiral order are needed to bring our results to better agreement
with the shape of the experimental spectra. The values 
for the combinations of LECs in Eqs.~(\ref{eq:ctcombeta}) and (\ref{eq:ctcombetap}) are
\beq
\begin{array}{lcr@{.}llcr@{.}l} \label{eq:par}
\bar{w}_{\eta}^{(m)}    & = & -2&21 \cdot 10^{-3} \,, \qq 
& \bar{w}_{\eta}^{(s)}  & = & -9&50  \cdot 10^{-3} \GeV^{-2} \,, \\
\bar{w}_{\eta'}^{(m)}   & = & -6&90  \cdot 10^{-3} \,, \qq 
& \bar{w}_{\eta'}^{(s)} & = &  1&00  \cdot 10^{-3} \GeV^{-2} \,. 
\end{array}
\eeq
It should be emphasized that the choice of parameters is not unique, since 
for $\eta, \eta' \to \pi^+ \pi^- \gamma$ variations in one 
of the parameters may be compensated by the other ones. However, 
with the choice in Eq.~(\ref{eq:par}) the remaining set of parameters is 
in agreement with previous work \cite{BN, BB3}.

In general, the values of the counter terms in the non-perturbative coupled channels approach
differ from those in the one-loop calculation as already observed for the two-photon 
decays \cite{BN} and the hadronic decays of $\eta$ and $\eta'$ \cite{BB3}, 
since in the loopwise expansion the effects of resonances are hidden in the LECs,
whereas they are generated dynamically in the non-perturbative approach.

\begin{figure}
\centering
\begin{minipage}{0.45\textwidth}
\includegraphics[width=1.0\textwidth,clip]{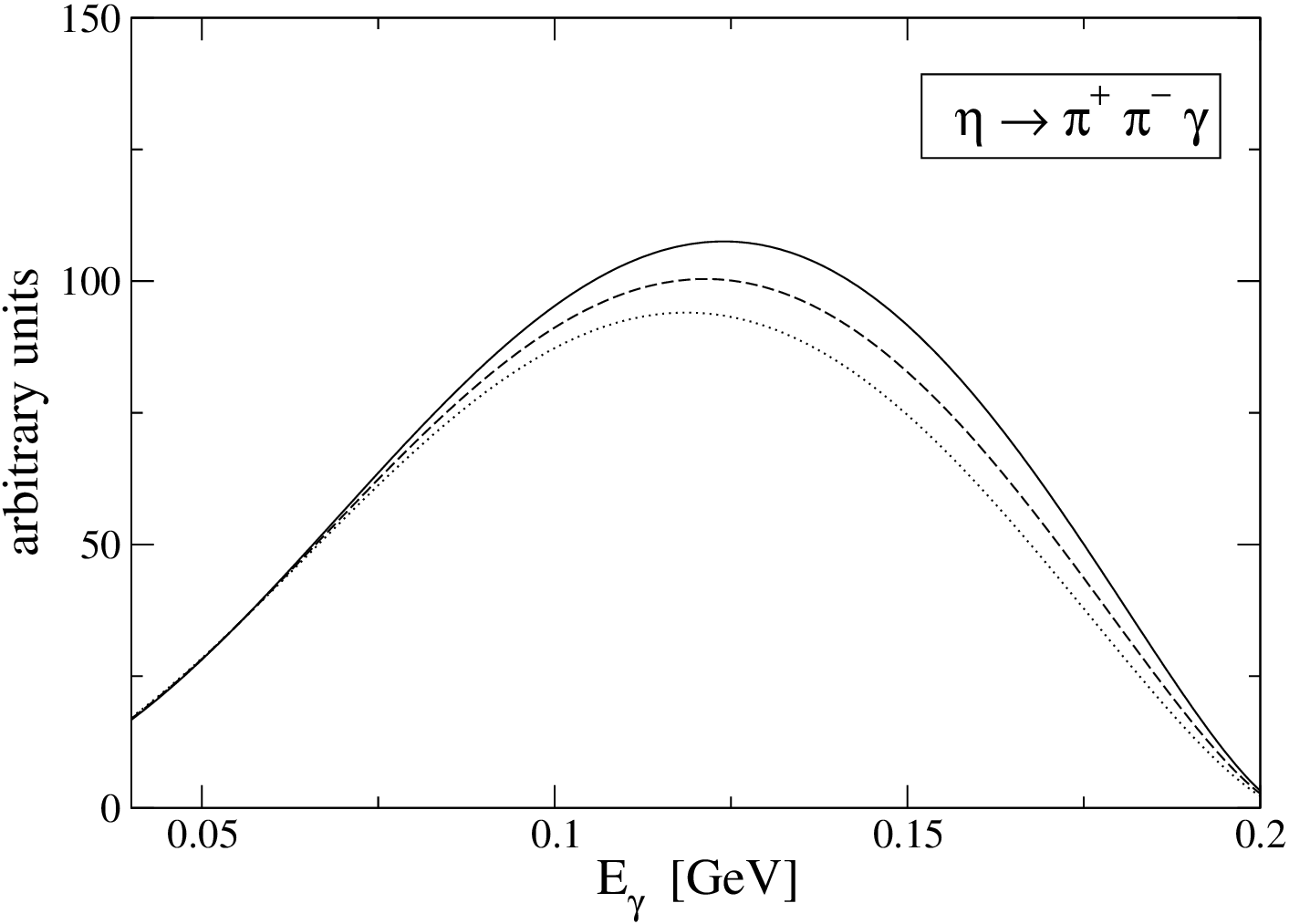}
\end{minipage}
\hspace{0.05\textwidth}
\begin{minipage}{0.45\textwidth}
\begin{overpic}[width=1.0\textwidth,clip]{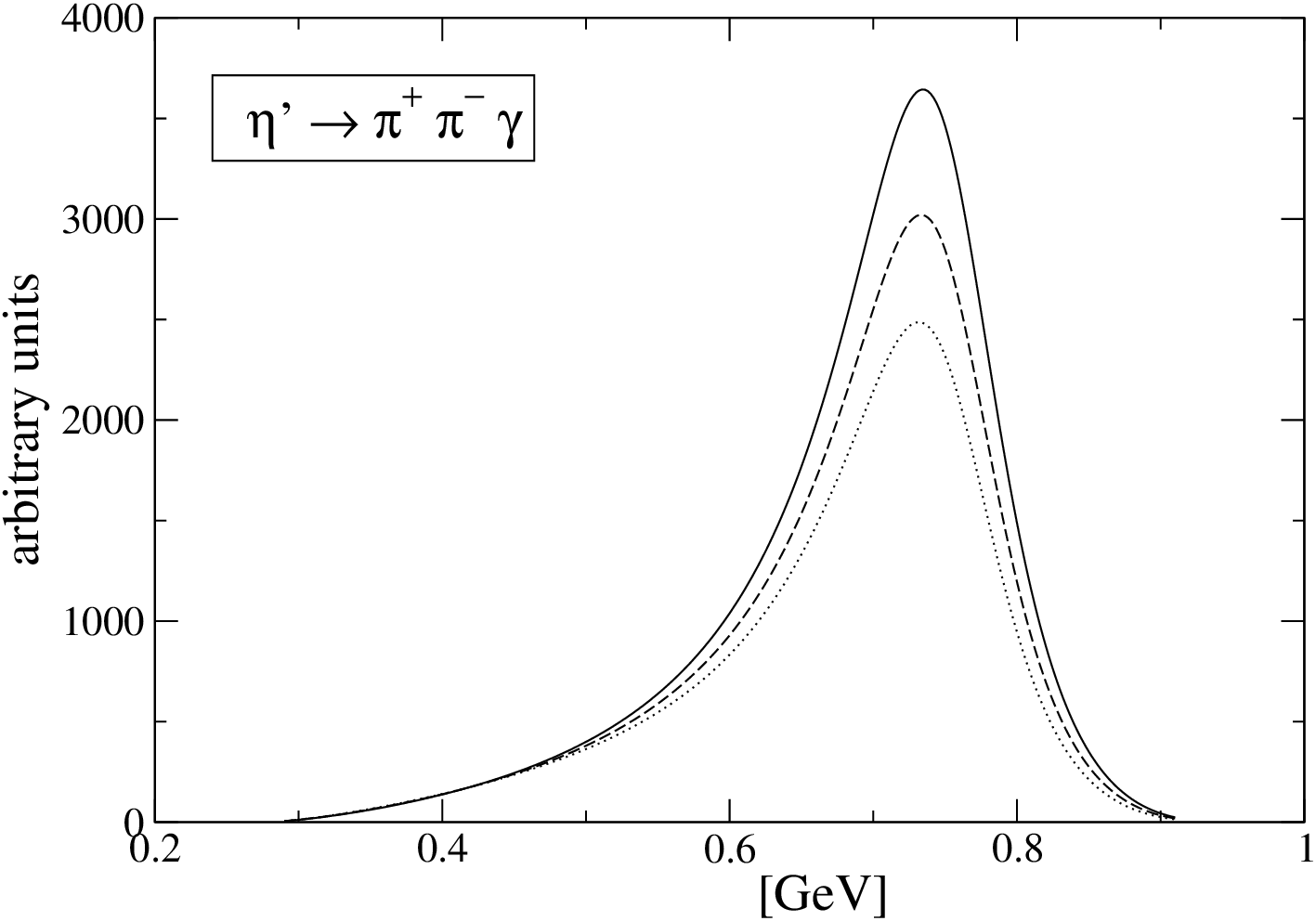}
\put(42,0.5){\scalebox{0.7}{$\sqrt{s_{+-}}$}}
\end{overpic}
\end{minipage}
\caption{Dependence on the mixing parameter $\cvtwid{2}{1}$:
 	 $\cvtwid{2}{1} = 0$ (solid), $\cvtwid{2}{1} = 0.6 F_{\pi}^{2}/4$ (dashed), 
 	 $\cvtwid{2}{1} = 1.2 F_{\pi}^{2}/4$ (dotted).}
\label{fig:v21t}
\end{figure}

In Fig.~\ref{fig:v21t} we show the dependence of our results on the $\eta$-$\eta'$ mixing 
parameter $\cvtwid{2}{1}$. In both decays the heights of the spectra are reduced for 
increasing values of $\cvtwid{2}{1}$ yielding smaller decay widths. As for the two-photon 
and hadronic decays, $\cvtwid{2}{1} \approx 0$ is thus in better agreement with the data.

\begin{figure}
\centering
\begin{minipage}{0.45\textwidth}
\begin{overpic}[width=1.0\textwidth,clip]{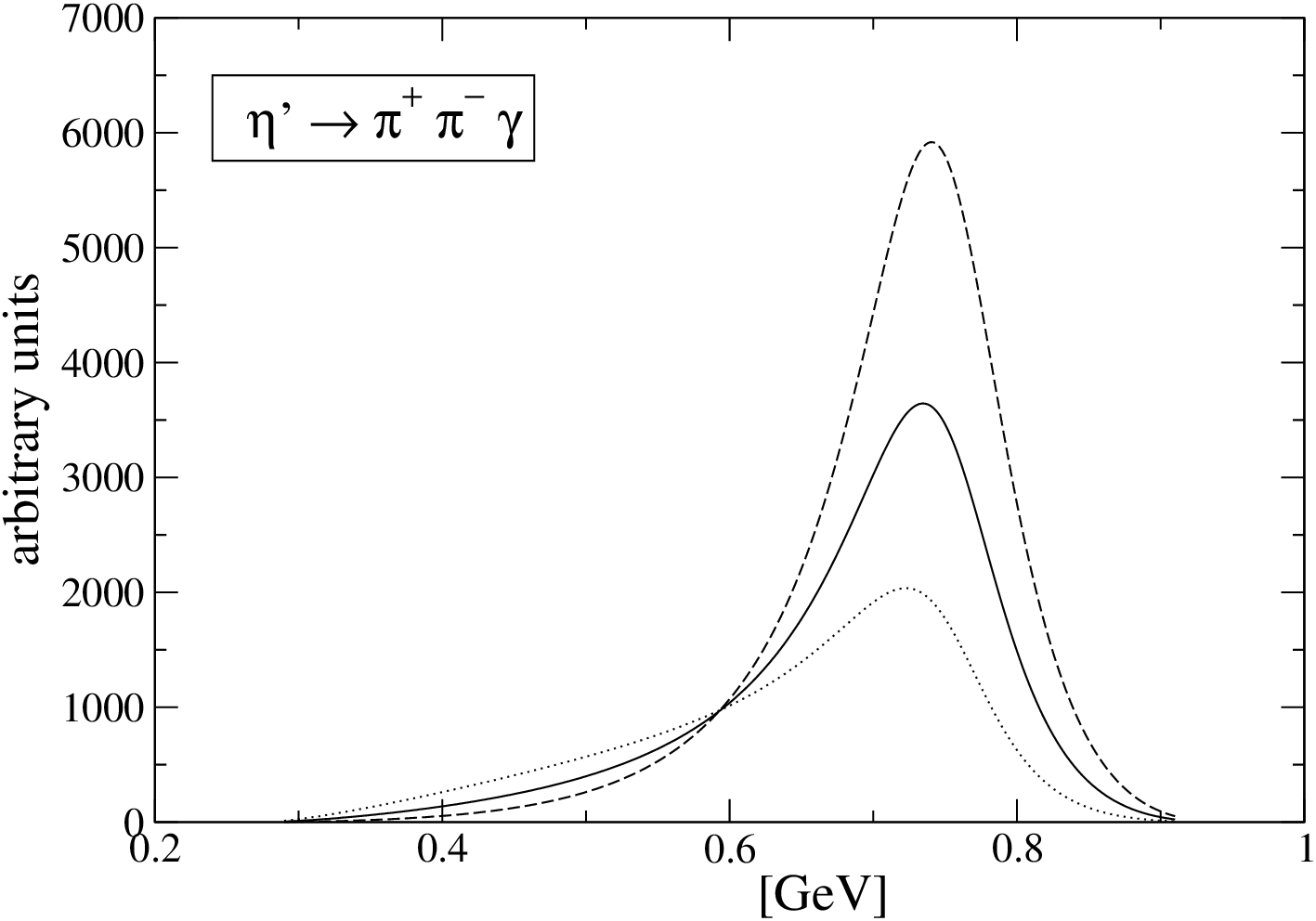}
\put(42,0.5){\scalebox{0.7}{$\sqrt{s_{+-}}$}}
\end{overpic}
\end{minipage}
\caption{Dependence on the LEC $w_{3}^{(1)r}$:
         $w_{3}^{(1)r} = -2.0 \times 10^{-3}$ (dashed), $w_{3}^{(1)r} = 0$ (solid), 
 	 $w_{3}^{(1)r} =  2.0 \times 10^{-3}$ (dotted).}
\label{fig:w31}
\end{figure}

The results for different values of the LEC $w_{3}^{(1)r}$ are depicted in 
Fig.~\ref{fig:w31}. When keeping $\cvtwid{2}{1} = 0$, the $\eta$ decay is not affected
significantly.
Similar to the two-photon decays negative values for $w_{3}^{(1)r}$ increase the $\rho$ peak
while positive values decrease it. Changes in $w_{3}^{(1)r}$ could in principle be compensated 
by altering the subtraction constant $C_\pi$ in the loop integral $\tilde{I}_1$.
The influence of $v_{1}^{(2)}$ on our results is rather small. Variations within a 
natural range for a coupling constant in the $\mathcal{O}(p^2)$ 
Lagrangian\footnote{The coefficients of 
$\trf{\partial_\mu U^\dagger \partial^\mu U}$ and $\trf{U^\dagger \chi + \chi^\dagger U}$ are
$f^2 / 4 \approx F_{\pi}^2 / 4 \approx 2.22 \cdot 10^{-3} \GeV^2$, for comparison.},
$-3.0 \cdot 10^{-3} \GeV^2, \dots, 3.0 \cdot 10^{-3} \GeV^2$, yield only slight corrections.

\begin{figure}
\centering
\begin{minipage}{0.45\textwidth}
\includegraphics[width=1.0\textwidth,clip]{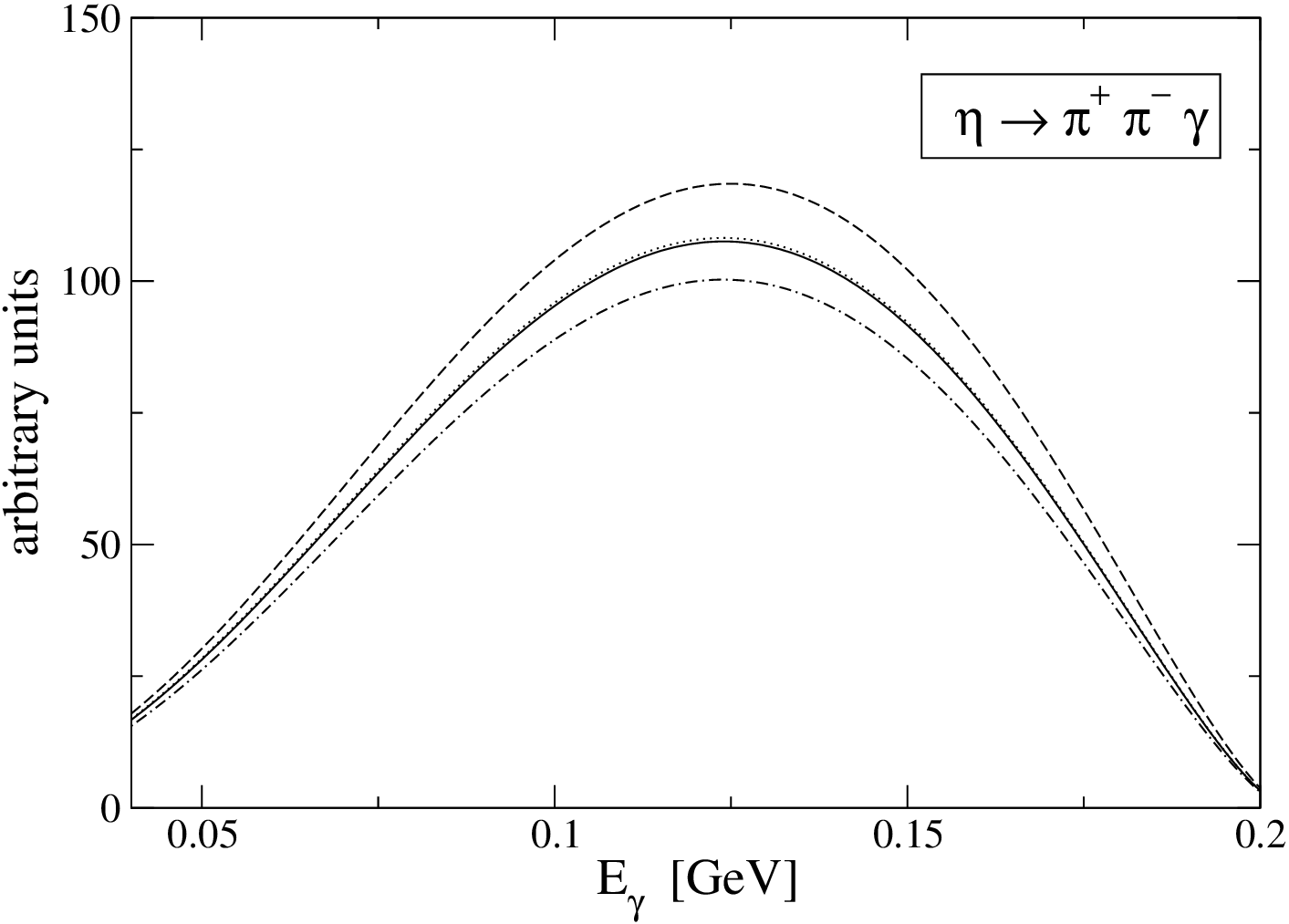}
\end{minipage}
\hspace{0.05\textwidth}
\begin{minipage}{0.45\textwidth}
\begin{overpic}[width=1.0\textwidth,clip]{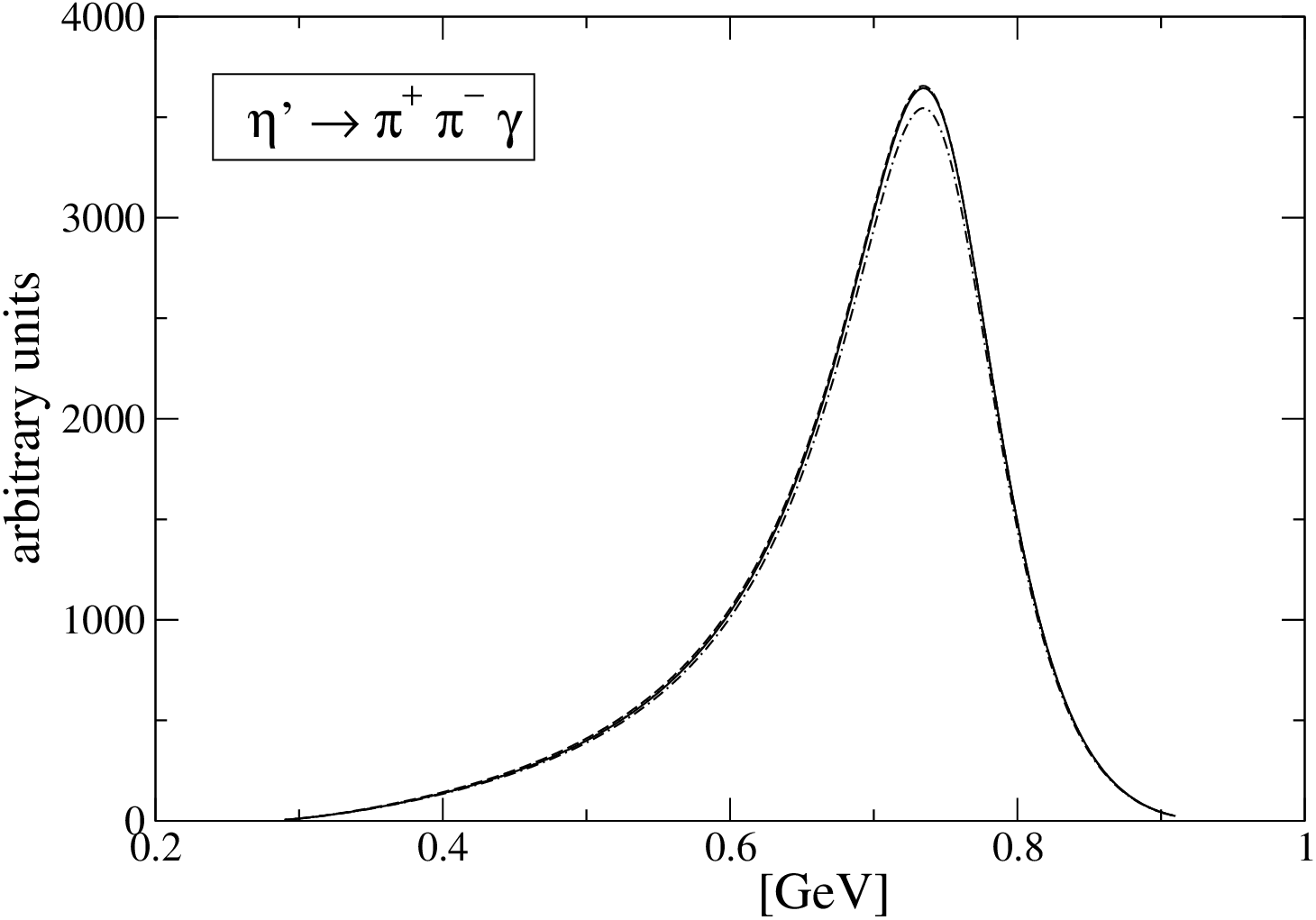}
\put(42,0.5){\scalebox{0.7}{$\sqrt{s_{+-}}$}}
\end{overpic}
\end{minipage} 
\caption{Comparison of the different coupled channel contributions in Fig.~\ref{fig:CC}:
         Full result (solid), result without $\mathcal{A}^{\textit(CCb)}$ (dashed),
	 without $\mathcal{A}^{\textit(CCc)}$ (dotted),
	 without $\mathcal{A}^{\textit(CCd)}$ (dot-dashed).}
\label{fig:CCabcd}
\end{figure}

Finally, we compare the contributions of the different coupled channel diagrams in 
Fig.~\ref{fig:CC}. From the plots in Fig.~\ref{fig:CCabcd} we see that the amplitude
for the pure $\pi^+ \pi^-$ final state interaction, $\mathcal{A}^{\textit{(CCa)}}$ 
(Eq.~(\ref{eq:ACCa})), furnishes by far the dominant part. The diagrams which involve the 
five-meson vertex from the WZW Lagrangian
(Fig.~\ref{fig:CC}b,~d) yield small corrections. This is in contradistinction to complete 
Vector Meson Dominance, where only these two diagrams are present. 
The contributions from the coupled channel diagram in Fig.~\ref{fig:CC}c are 
almost negligible.


\section{Conclusions} \label{sec:concl}

In this work, we have calculated the anomalous decays
$\eta , \eta' \to \pi^+ \pi^- \gamma$ within an approach that combines
ChPT with a non-perturbative scheme based on coupled channels. This method
satisfies unitarity constraints and generates vector mesons from composed states
of pseudoscalar mesons without including them explicitly in the effective Lagrangian.
It had recently been applied in the anomalous sector
for the two-photon decays of $\pi^0, \eta$ and $\eta'$ and is now extended to 
the decays into $2 \pi \gamma$.

We first performed a full one-loop calculation within the framework of
chiral perturbation theory and without imposing large $N_c$ counting rules. 
The presence of the massive state $\eta'$ spoils the strict chiral counting scheme such that
loop diagrams with an $\eta'$ also contribute at fourth chiral order, {\it i.e.}\ 
at the same order as tree level contributions from the WZW Lagrangian.
We have shown explicitly that these contributions can be absorbed into a non-anomalous
contact interaction of unnatural parity, but they do not renormalize the WZW term
so that the anomalous Ward identities are preserved. 

If the counter terms of unnatural parity at subleading chiral order $p^6$
are omitted in the one-loop calculation, 
the results are in disagreement with the experimental decay widths.
Taking $\eta$-$\eta'$ mixing into account ameliorates the situation slightly for the $\eta$ decay,
but is still in contradiction to experiment.
As a matter of fact, it is not possible to constrain $\eta$-$\eta'$ mixing from
the decay $\eta \to \pi^+ \pi^- \gamma$. 
Moreover, one should keep in mind that mixing effects are of sixth chiral order,
if large $N_c$ counting rules are not employed, and enter at the same order
as the neglected counter terms.
The inclusion of contact interactions of sixth chiral order improves the situation
for the $\eta$ decay, whereas the results for the $\eta'$ decay cannot be brought to agreement 
with experiment by adjusting the counter terms of sixth chiral order. 
This is due to the presence of vector mesons, which dominate the $\eta'$ decay.
While for the $\eta$ decay only effects of the tail of the resonances
contribute which can be treated perturbatively by absorbing them into the couplings
of the effective Lagrangian, one must include unitarity effects via final state
interactions in the $\eta'$ decay.

The inclusion of the coupled channel formalism provides a framework which
reproduces both the decay widths and the experimental dipion invariant-mass spectra 
of the $\eta$ and $\eta'$ decays
while matching onto the results of ChPT. Our choice of parameters is consistent with that
of the two-photon decays as discussed in \cite{BN} and provides a non-trivial check
for our approach. It furthermore confirms that the inclusion of
unitarity effects via the Bethe-Salpeter equation can be accomplished in the
anomalous sector of QCD in a similar way as 
for the hadronic decays of $\eta$ and $\eta'$ \cite{BB3}.
Finally, we have shown that this method does not renormalize the WZW term
and hence satisfies constraints from anomalous Ward identities.

\section{Acknowledgements} \label{sec:ackn}

We would like to thank Edisher Lipartia for useful discussions.

\begin{appendix}

\section{$\mathcal{O}(p^6)$ Contact Terms in $P \to \pi^+ \pi^- \gamma$} 
  \label{app:ct}

In this section we discuss the $\mathcal{O}(p^6)$ counter terms which contribute to  
$\eta$ and $\eta'$ decays into $\pi^+ \pi^- \gamma$. We use the notation of \cite{BN}
for the covariant derivative $D_\mu U$ and the field strength tensors $\tilde{R}_{\mu \nu}$ 
and $\tilde{L}_{\mu \nu}$ of the right- and left-handed external fields, respectively,
and make use of the following 
abbreviations:
\beq \label{eq:ctabbrev}
\begin{array}{lcllcl}
\tilde{P}_{\mu \nu} & = & U^{\dagger} \tilde{R}_{\mu \nu} U + \tilde{L}_{\mu \nu} \,, \qq &
\tilde{Q}_{\mu \nu} & = & U^{\dagger} \tilde{R}_{\mu \nu} U - \tilde{L}_{\mu \nu} \,, \\
M & = & U^\dagger \chi+\chi^\dagger U \,, \qq &
N & = & U^\dagger \chi-\chi^\dagger U \,, \\
C_{\mu} & = & U^\dagger \cder_\mu U \,, \qq &
E_{\mu \nu} & = & U^\dagger D_\mu D_\nu U - (D_\mu D_\nu U)^\dagger U \ .
\end{array}
\eeq
These are the building blocks for the construction of the counter terms.
Essentially, there are two types of counter terms; either they include the mass matrix 
$\mathcal{M} = \mbox{diag}(\hat{m},\hat{m},m_s)$ 
or they contain two additional derivatives 
resulting in contributions to the decay amplitude proportional to $s_{+ -}$ which can be expressed
in terms of the photon energy
$\omega$. In the $SU(3)$ framework the entire set of  $\mathcal{O}(p^6)$ terms of unnatural parity
has been presented in \cite{ChPTO6}, but the inclusion of the singlet field $\eta_0$ 
induces additional structures. The mass terms may be written as
\beqa
\Lagr_{\chi}^{(6)} & = & \epsilon^{\mu \nu \alpha \beta} \Big\{ 
  \bar{W}_7 \trf{N (\tilde{P}_{\mu \nu} C_\alpha C_\beta + C_\alpha C_\beta \tilde{P}_{\mu \nu}
                 + 2 C_\alpha \tilde{P}_{\mu \nu} C_\beta)} \no \\
& & \qq \q + \bar{W}_8 \left(\trf{M C_\mu} \trf{C_\nu \tilde{Q}_{\alpha \beta}} 
                  + \trf{N} \trf{\tilde{P}_{\mu \nu} C_\alpha C_\beta} \right) \no \\
& & \qq \q + \bar{W}_9 \left(\trf{M (\tilde{Q}_{\mu \nu} C_\alpha 
                             + C_\alpha \tilde{Q}_{\mu \nu})}
                  + \trf{N (\tilde{P}_{\mu \nu} C_\alpha - C_\alpha \tilde{P}_{\mu \nu})}
                  \right) \trf{C_\beta} \no \\
& & \qq \q + \bar{W}_{10} \trf{M} \trf{\tilde{Q}_{\mu \nu} C_\alpha} \trf{C_\beta} \Big\} \,,
\eeqa
where the last two terms arise from the extension to $U(3)$.
Expansion in the meson fields yields
in the differential form notation of \cite{KL1}
{\arraycolsep3pt
\beqa \label{eq:ctm}
d^4x \mathcal{L}_{\chi, \,ct}^{(6)} & = &
    i\,\bar{w}_{7}^{(0)} \,\frac{8 \sqrt{2}}{f^3} 
      \trf{\{\chi \,,\phi\} (\{d\phi \,d\phi \,, dv\} + 2 d\phi \,dv \,d\phi)} \no \\
& & +\, i\,\bar{w}_{8}^{(0)} \,\frac{32 \sqrt{2}}{f^3} \trf{\chi \,\phi}
      \trf{d\phi \,d\phi \,dv} 
    -\, i\,\bar{w}_{9}^{(0)} \,\frac{16 \sqrt{6}}{f^3} \,\eta_0 
      \trf{\{\chi \,,d\phi\}[d\phi \,,dv]}
\no \\
& & -\, i\,\bar{w}_{10}^{(0)}\,\frac{16 \sqrt{6}}{f^3} \,\eta_0 
      \trf{\chi} \trf{d\phi \,d\phi \,dv} 
\eeqa}%
and we have integrated by parts in order to simplify the structures.
The terms with five derivatives and one external field are
\beqa
\Lagr_{\partial}^{(6)} & = & \epsilon^{\mu \nu \alpha \beta} \Big\{
      \bar{W}_{11} \trf{\tilde{P}_{\mu \nu} (E^{\lambda}_{\ \alpha} C_\beta C_\lambda 
    - C_\lambda C_\beta E^{\lambda}_{\ \alpha})} \no \\
& & + \bar{W}_{12} \trf{\tilde{P}_{\mu \nu} (E^{\lambda}_{\ \alpha} C_\lambda C_\beta
    - C_\beta C_\lambda E^{\lambda}_{\ \alpha})} \no \\
& & + \bar{W}_{13} \trf{\tilde{P}_{\mu \nu} (E^{\lambda}_{\ \alpha} C_\lambda
    - C_\lambda E^{\lambda}_{\ \alpha})} \trf{C_\beta} \no \\
& & + \bar{W}_{14} \trf{\tilde{P}_{\mu \nu} (E^{\lambda}_{\ \alpha} C_\beta
    - C_\beta E^{\lambda}_{\ \alpha})} \trf{C_\lambda} \Big\} 
\eeqa
with $\bar{W}_{13}$ and $\bar{W}_{14}$ being only present in the $U(3)$ framework. Expansion in $\phi$ 
yields the vertices
\beqa \label{eq:ctp}
d^4x \Lagr_{\partial, \,ct}^{(6)} & = & 
    -\,i \,\bar{w}_{11}^{(0)} \,\frac{16 \sqrt{2}}{f^3} 
       \trf{(\partial^\lambda d\phi \,d\phi \,\partial_\lambda \phi 
            + \partial_\lambda \phi \,d\phi \,\partial^\lambda d\phi) \,dv} \no \\
& & -\,i \,\bar{w}_{12}^{(0)} \,\frac{16 \sqrt{2}}{f^3}
       \trf{(\partial^\lambda d\phi \,\partial_\lambda \phi \,d\phi
            + d\phi \,\partial_\lambda \phi \,\partial^\lambda d\phi) \,dv} \no \\
& & -\,i \,\bar{w}_{13}^{(0)} \,\frac{16 \sqrt{2}}{f^3} \trf{d\phi}
       \trf{[\partial^\lambda d\phi \,, \partial_\lambda \phi] \,dv} \no \\
& & -\,i \,\bar{w}_{14}^{(0)} \,\frac{16 \sqrt{2}}{f^3} \trf{\partial_\lambda \phi}
       \trf{[\partial^\lambda d\phi \,, d\phi] \,dv} \ .
\eeqa

\end{appendix}


\end{document}